\documentclass[
reprint,
 amsmath,amssymb,
 aps,
 prl,
]{revtex4-2}
\UseRawInputEncoding
\usepackage{graphicx}
\usepackage{dcolumn}
\usepackage{bm}
\usepackage[mathlines]{lineno}

\usepackage{xcolor}

\begin{document}


\title{Geometry of optimal control in chemical reaction networks}

\author{Yikuan Zhang}
\affiliation{%
School of Physics, Peking University, Beijing 100871, China 
}%
\author{Qi Ouyang}%
\affiliation{%
Institute for Advanced Study in Physics, Zhejiang University, Hangzhou 310058, China
}%
\affiliation{%
Center for Quantitative Biology,
AAIC, Peking University, Beijing 100871, China
}%
\author{Yuhai Tu}%
\affiliation{IBM T. J. Watson Research Center,
Yorktown Heights, New York 10598, USA}


\begin{abstract}
Although optimal control (OC) has been studied in stochastic thermodynamics for systems with continuous state variables, less is known in systems with discrete state variables, such as Chemical Reaction Networks (CRNs). Here, we develop a general theoretical framework to study OC of CRNs for changing the system from an initial distribution of states to a final distribution with minimum dissipation.   
We derive a ``Kirchhoff's law" for the probability current in the adiabatic limit, from which the optimal kinetic rates are determined analytically for any given probability trajectory. By using the optimal rates, we show that the total dissipation is determined by a $L_2$-distance measure in the probability space 
and derive an analytical expression for the metric tensor that depends on the probability distribution, network topology, and capacity of each link. Minimizing the total dissipation leads to the geodesic trajectory in the probability space and the corresponding OC protocol is determined by the Kirchhoff's law.  To demonstrate our general approach, we use it to find a lower bound for the minimum dissipation that is tighter than existing bounds obtained with only global constraints. We also apply it to simple networks, e.g., fully connected 3-state CRNs with different local constraints and show that indirect pathway and non-functional transient state can play a crucial role in switching between different probability distributions efficiently. Future directions in studying OC in CRNs by using our general framework are discussed. 
\end{abstract}

\maketitle

Most biological systems operate out of thermal equilibrium by constantly dissipating energy and exchanging matter and information with their environment. Despite their high noise level, biological systems can control (regulate) the underlying biochemical networks to achieve different biological functions accurately driven by chemical energy dissipation from 
hydrolysis of high energy molecules such as ATP and GTP~ \cite{Ken_energetics,Udo_lambda,Hopfield_KPR,Hill_transduction,Udo_CRNs,Esposito_CRNs,Qian_Phosphorylation,Tu_oscillations,Tu_chemotaxis,Tu_synchronization,Udo_TUR,Udo_cost,Udo_PEC,Howad_oscillations}. 
As control theory gains attention in biology in recent years~\cite{control1,control2,control3,ilker_2022}, an important general problem is to identify the optimal control process (protocol) to achieve certain biological function with minimal energy dissipation, i.e., the optimal control (OC) problem.

Within the fields of finite-time thermodynamics and stochastic thermodynamics, much work has been done in finding the optimal protocol that leads to the minimum total entropy production during thermodynamic processes \cite{OC_2007,OC_2008,LP_2020,OC_2015,OC_2017,OC_2021,Udo_CRN_2021,OC_2022_2,OC_2023,OT_2011,OT_2021,OT_2023_2,Dechant_dec1,Geo_1975,Geo_1983,Geo_1988,Geo_2007,Geo_2012,Geo_2020,Geo_2022} especially in systems with continuous state variables that can be described by overdamped Langevin dynamics \cite{OC_2007,OC_2008,LP_2020,OC_2022_2,OC_2023,OT_2011,OT_2021,OT_2023_2,Dechant_dec1}.

However, for systems with discrete state variables such as CRNs, most of the existing work focused on establishing lower bounds of the total entropy production ~\cite{SL_2018,Van_2021,CRNs_2021,OT_2022,OT_2023,Dechant_dec2} without considering OC. For example, Vu et al. computed dissipation in systems with given time-dependent transition rates that satisfy instantaneous detailed balance condition, yielding a lower bound without considering the OC problem~\cite{Van_2021}.  
By using the Wasserstein distance on graphs, Dechant obtained a lower bound with the global constraint of fixed total reaction activity~\cite{OT_2022}. 
Muratore-Ginanneschi et al, did consider the OC problem but only for a continuous-time Markov jump process in one-dimensional countable state space~\cite{DO_2013}.

Given that most biological systems are governed by biochemical networks with discrete state variables,  
we aim to develop a general theoretical framework to study the OC problem in CRNs under realistic constrains in this paper.

$\textbf{\textit{The OC problem in discrete systems}}$. In a CRN with $N$ states, the transitions among these states are described by the Chemical Master equation (CME):
\begin{equation}\label{eq1}
\frac{d P_i}{d t}=\sum_j\left(k_{j i} P_j-k_{i j} P_i\right), \quad i=1,2,3, \ldots, N,
\end{equation}
where $P_i(t)$ is the probability in state-$i$ at time $t$ and $k_{ij}$ is the transition
probability rate from state-$i$ to state-$j$ .

The goal of the control process is to change the probability distribution of the system from $\textbf{P}(0)$ to a final distribution $\textbf{P}(\tau)$ by changing the $(N\times N)$ transition rate matrix $\mathbf{K}(t)$ during $0 \leq t \leq \tau$. During this control process, the total energy dissipation is given by the total entropy production
($k_BT=1$): 
\begin{equation}\label{eq2}
\mathcal{C}(\textbf{K}(t))=\int_{0}^{\tau} \dot{S}(\mathbf{K}(t), \textbf{P}(t)) dt,
\end{equation}
where $\dot{S}(t)=\frac{1}{2}\sum_{i j}\left(k_{i j} P_i-k_{j i} P_j\right) \ln \frac{k_{i j} P_i}{k_{j i} P_j}$ is the dissipation rate at time $t$. 

The optimal control problem is to find the optimal protocol $\mathbf{K}(t)$ that minimizes the cost $\mathcal{C}(\textbf{K}(t))$ for changing the probability distribution from an initial probability distribution $\textbf{P}(0)$ for $t\le 0$ to the final distribution $\textbf{P}(\tau)$ for $t\ge \tau$. 
Following previous work~\cite{OC_2007,OC_2021}, we allow rates to have discontinuous jumps at $t=0$ and $t=\tau$.

Since each bi-directional reaction has different reactants and can be controlled by different enzymes, its reaction rates ($k_{ij}$ and $k_{ji}$) are subject to reaction-specific constraint.  For simplicity, we impose a reaction-specific constraint: 
$k_{ij} + k_{ji}=C_{ij}$ with $C_{ij}$ a constant local rate capacity for link-$ij$. Other local constraints can be used without affecting the general results (see the 3-state gene circuit example for another choice of local constraints).

An alternative is to constraint the time averaged global reaction activities  $\bar{A}:=\frac{1}{\tau} \int_0^\tau d t A(t), A = \sum_{i>j}P_ik_{ij}$ \cite{SL_2018,OT_2022,OT_2023}, which is less realistic given that individual reactions may be subject to different constraints.

$\textbf{\textit{The two-step optimization scheme}}.$ In systems with continuous state variables, OC 
is a difficult problem involving solving a nonlocal Euler-Lagrange equation in the joint rate-probability space~\cite{OC_2007}. 
For the discrete systems studied here, we adopt a two-step optimization scheme, which makes the problem tractable. 

The first step is to find the optimal $\textbf{K}^{*}(t)$ for any given probability trajectory $\textbf{P}(t)$ similar to the approach used by Ilker et al~\cite{ilker_2022}. For a given $\textbf{P}(t)$, the CME (Eq.~\ref{eq1}) can be considered as a set of linear equations for $\textbf{K}(t)$. Since the number of control variables (i.e. the number of $k_{ij}$) is larger than $N$, there are many solutions for $\textbf{K}(t)$, and the first step of optimization is to find the optimal rates $\textbf{K}^{*}(\textbf{P}(t),\dot{\textbf{P}}(t))$ that minimizes the entropy production $\mathcal{C}(\textbf{K}(t))$. 

The second step of optimization is then to find the optimal $\textbf{P}^{*}(t)$ trajectory for given initial and final distributions $\textbf{P}(0)$, $\textbf{P}(\tau)$ to minimize the entropy production: 
\begin{equation}\label{varP}
\begin{gathered}
\textbf{P}^{*}(t) = \mathop{\arg\min}\limits_{\textbf{P}}\int_{0}^{\tau} \dot{S}(\mathbf{K}^{*}(\textbf{P}(t),\dot{\textbf{P}}(t)), \textbf{P}(t)) d t,
\end{gathered}
\end{equation}
which reduces to a variational problem w.r.t. $\textbf{P}(t)$. Solving the resulting Euler-Lagrange equation for $\textbf{P}(t)$ leads to the optimal probability trajectory $\textbf{P}^{*}(t)$.

\begin{figure}
\includegraphics[scale=0.57]{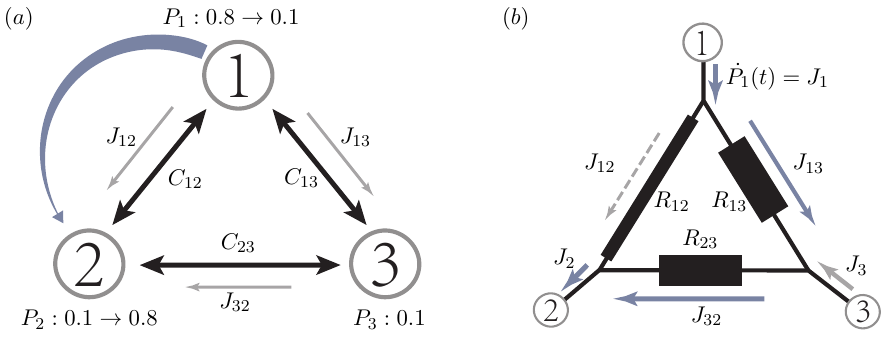}
\caption{\label{fig1} The circuit analogy of CRN. (a) The optimal control problem of the smallest CRN, a fully connected 3-state network. (b) The corresponding electronic circuit.}
\label{illustration}
\end{figure}

$\textbf{\textit{Analogy to electronic circuit}}$.  
In the adiabatic limit where the relaxation time of the system $\tau_R$ is much shorter than $\tau$,  
we have $|1-P_ik_{ij}/P_jk_{ji}|=\mathcal{O}(\tau_R/\tau)\ll 1$, which means that the system is near equilibrium. The total dissipation can then be approximated as:
\begin{equation}\label{eq5}
\begin{aligned}
  \mathcal{C}(\mathbf{K})\approx \int_{0}^{\tau}\sum_{ij}\frac{(P_ik_{ij}-P_jk_{ji})^2}{P_ik_{ij}+P_jk_{ji}} dt=\int_{0}^{\tau}\sum_{ij}J_{ij}^2R_{ij} dt,
\end{aligned}
\end{equation}
which is analogous to the Joule's law for electronic circuit. As shown in Fig.~\ref{fig1}(b),  $J_{ij}=-J_{ji}=P_ik_{ij}-P_jk_{ji}$ is the net current from node-$i$ to node-$j$; $R_{ij}=R_{ji}=(P_ik_{ij}+P_jk_{ji})^{-1} \approx \frac{1}{2C_{ij}}(\frac{1}{P_i(t)}+\frac{1}{P_j(t)})$ is the resistance for link-$ij$~\footnote{In the adiabatic limit, the control process is carried out near-equilibrium: $P_ik_{ij}\approx P_jk_{ji} (1+\mathcal{O}(\tau_R/\tau))$. Together with the constraint $k_{ij}+k_{ji}=C_{ij}$, we have $R_{ij}\approx \frac{1}{2C_{ij}}(\frac{1}{P_i(t)}+\frac{1}{P_j(t)})$ to the leading order in $\mathcal{O}(\tau_R/\tau)$.}; $P_i$ 
is the charge at node-$i$. Since $P_i$ depends on time (for changing probability), there is an input (driving) current source $J_i(t)=\dot{P}_i(t)$ at each node-$i$ with the constraint $\sum_i J_i=0$ due to charge conservation.  

From the circuit analogy, the first step in solving the OC problem becomes optimizing $\mathcal{C}$ in Eq.~\ref{eq5} with the current conservation constraint $J_i(t)=\Sigma_{j}J_{ij}(t)$ at each node-$i$. By introducing a Lagrange multiplier $V_i(t)$ analogous to the voltage at node-$i$, we can solve this optimization problem in the functional space of $J_{ij}$ to obtain: 
\begin{equation}\label{kirchhoff}
\sum_j J_{ij}=\sum_j (V_i-V_j)/R_{ij}=J_i(t)=\dot{P}_i(t),\;\;\forall i
\end{equation}
which is the Kirchhoff's law for the CRN circuit. Note that the circuit analogy for CRNs has been investigated in recent studies~\cite{Circuit_2020,Circuit_2023}, however, as far as we know it has not been used in the context of OC.

By introducing the conductance matrix $\bf{\Gamma}(\textbf{R})$ 
with $\Gamma_{ii}=\sum_j R_{ij}^{-1}$ and $\Gamma_{ij}=-R_{ij}^{-1}$ ($i\ne j$), Eq.~\ref{kirchhoff} can be solved (formally):  $\textbf{V}=\bf{\Omega}\dot{\textbf{P}}$ with $\bf{\Omega}\equiv \bf{\Gamma}^{-1}$~\footnote{The inverse here is formal, since $\bf{\Gamma}(\textbf{R})$ is a singular matrix due to $\sum_iJ_i(t)=\sum_i\dot{P}_i(t)=0$, see Sec.~I.A in SM for the definition of the inverse}. The optimal current for a given $\textbf{P}(t)$ can then be obtained: $J_{ij}=\sum_k (\Omega_{ik}-\Omega_{jk})R^{-1}_{ij} \dot{P}_k\equiv \sum_k \Phi_{ij,k}(\textbf{R})\dot{P}_k$ where $\Phi_{ij,k}(\textbf{R})= (\Omega_{ik}-\Omega_{jk})R^{-1}_{ij}$ is a $(L\times N)$ matrix with $L$ the number of transitions in the network.
From the optimal current, we can determine the optimal control protocol of the rates: $k_{ij}^*=(P_i+P_j)^{-1} (P_j C_{ij}+\sum_{k}\Phi_{ij,k}\dot{P}_k)$ for any given $\mathbf{P}(t)$~\footnote{In the expression for $k_{ij}^*=(P_i+P_j)^{-1} (P_j C_{ij}+\sum_{k}\Phi_{ij,k}\dot{P}_k)$, the first term $(P_i+P_j)^{-1} P_j C_{ij}$ represents the equilibrium effect when detailed balance is satisfied, and the second term $(P_i+P_j)^{-1} \sum_{k}\Phi_{ij,k}\dot{P}_k$, which is of order $\mathcal{O}(\tau_R/\tau)$, represents the out-of-equilibrium effect that breaks detailed balance.}.

\textit{\bf Dissipation length and geodesic trajectory in probability space}.  
By using the optimal control protocol for given probability trajectory $\textbf{P}(t)$, we obtain an analytical expression for the total dissipation rate in the probability space with a quadratic form in $\dot{\textbf{P}}$:
\begin{equation}\label{geo_l}
    \dot{S}=\sum_{ij} \Lambda_{ij}(\textbf{R})\dot{P}_i\dot{P}_j,
\end{equation}
where $\Lambda_{ij}(\textbf{R})\equiv \Sigma_{kl}  \Phi(\textbf{R})_{kl,i}\Phi(\textbf{R})_{kl,j}R_{kl}$ is defined as the ``dissipation metric", which 
depends on $\textbf{P}$ and the capacity matrix $\textbf{C}$ through its direct dependence on $\mathbf{R}$. 

The dissipation rate given by Eq.~\ref{geo_l} transforms the original OC problem defined in the $\mathbb{R}_+^L$ rate-space to an optimal transport problem in the smaller $\mathbb{R}_+^{N-1}$ probability-space. The dissipation metric can be used to define a ``dissipation length" in the probability space:
\begin{equation}
    \mathcal{L}\equiv \int_0^{\tau} \sqrt{\sum_{i,j=1}^{N-1} \Lambda^*_{ij}(\textbf{R})\dot{P}_i\dot{P}_j}dt,
\end{equation}
where $\bf{\Lambda}^*$ is the reduced dissipation metric by taking into account  the constraint $\sum_i^N P_i=1$ (see Sec.~I in SM for details). By using the Cauchy-Shwartz inequality, the minimum dissipation is obtained: 
\begin{equation}
   min(\mathcal{C})=\mathcal{L}^2_{m}/\tau,
\end{equation} 
where $\mathcal{L}_{m}$ is the minimum dissipation length of the optimal probability trajectory $\textbf{P}^*(t)$, which can be determined by solving the geodesic equation (see Sec.~I in SM for details). The corresponding OC protocol can be obtained subsequently from the dependence of $\textbf{K}^*(t)$ on $\textbf{P}^*(t)$ given by the Kirchhoff's law.  

$\textbf{\textit{The Kirchhoff bound}} $. To demonstrate our approach, we first used it to establish a lower bound for the total entropy production. In the adiabatic limit, we have:
$R_{ij}\approx \frac{1}{2C_{ij}}(\frac{1}{P_i(t)}+\frac{1}{P_j(t)})\ge \frac{2}{C_{ij}(P_i+P_j)}>2/C_{ij}$. If we replace $R_{ij}$ by its lower bound $2/C_{ij}$, and minimize the resulting cost $\tilde{\mathcal{C}}\equiv \int_{0}^{\tau}2\sum_{ij}\frac{J_{ij}^2}{C_{ij}} dt$, we can obtain a lower bound for the total entropy production $\mathcal{C}$. Specifically, since $R_{ij}$ is replaced by a constant $2/C_{ij}$, the metric matrix $\bf{\Lambda}$ becomes independent of $\textbf{P}$ corresponding to a flat manifold. Thus, solving the geodesic equation leads to a linear probability trajectory $P_i(t)=P_i(0)+\Delta P_i \frac{t}{\tau}$, where $\Delta P_i = P_i(\tau)-P_i(0)$. The corresponding $\tilde{\mathcal{C}}$ sets a lower bound for the total entropy production, which we call the Kirchhoff bound:

\begin{equation}\label{Newbound}
\begin{aligned}
    \Sigma_{kh} :&= \sum_{ij} \int_{0}^{\tau}\Lambda_{ij}(\textbf{C})\dot{P}_i\dot{P}_j dt= \Delta \textbf{P}^T \textbf{R}^{*} \Delta \textbf{P}/\tau,
\end{aligned}
\end{equation}
where $R^*_{ij}$ is the effective resistance that contains the topological information of the CRN and strengths of its links.

The Kirchhoff bound (Eq.~\ref{Newbound}) is generally applicable to any CRN. To understand its physical meaning, we computed the Kirchhoff bound explicitly for the transport process of a fully connected 3-state network  (Fig.~\ref{illustration}(a)): 
\begin{equation}
    \Sigma_{kh}=2\frac{ \lVert \textbf{P}(0)-\textbf{P}(\tau)\rVert^2}{(C_{12} + C_{13} C_{23}/(C_{13}+C_{23}) )\tau},
\end{equation}
which has an intuitive interpretation: the numerator quantifies the amount of probability that needs to be transported, and the denominator is the effective conductance of the network.

We have compared the Kirchhoff bound with existing lower bounds~\cite{OT_2022,OT_2023} obtained based on Wasserstein distance on graph $\mathcal{W}_1\left(p(0), p(\tau)\right)$ with a global constraint by getting rid of the $P(t)$ dependence of this constraint using the same inequality to $R_{ij}(t)$ . 
We found that the Kirchhoff bound $\Sigma_{kh}$ is closer to the exact numerical value of the minimum entropy production than the Wasserstein bound $\Sigma_{wst}$ (see Sec.~II in SM for details). Intuitively, $\Sigma_{wst}$ results from connecting all the channels 1-2, 1-3 and 3-2 in parallel regardless of the structure of the network, whereas $\Sigma_{kh}$ is derived from the Kirchhoff's law that contains the topological information of the CRN, which makes $\Sigma_{kh}$ a tighter bound.

$\textbf{\textit{Modes of optimal probability transport}}$.  Next, we applied our approach to study the OC problem in simple networks such as the 2-state model, which can be solved analytically, see Sec.~I.C in SM for details. Here, we study the optimal control of changing the distribution
from predominately in state-1 (P1 = 0.8, P2 = P3 = 0.1)
at t = 0 to predominantly in state-2 (P2 = 0.8, P1 =
P3 = 0.1) in the fully connected 3-state model (Fig.~\ref{fig1}(a)). 
\begin{figure}[!ht]
	\centering
	\includegraphics[scale=0.4]{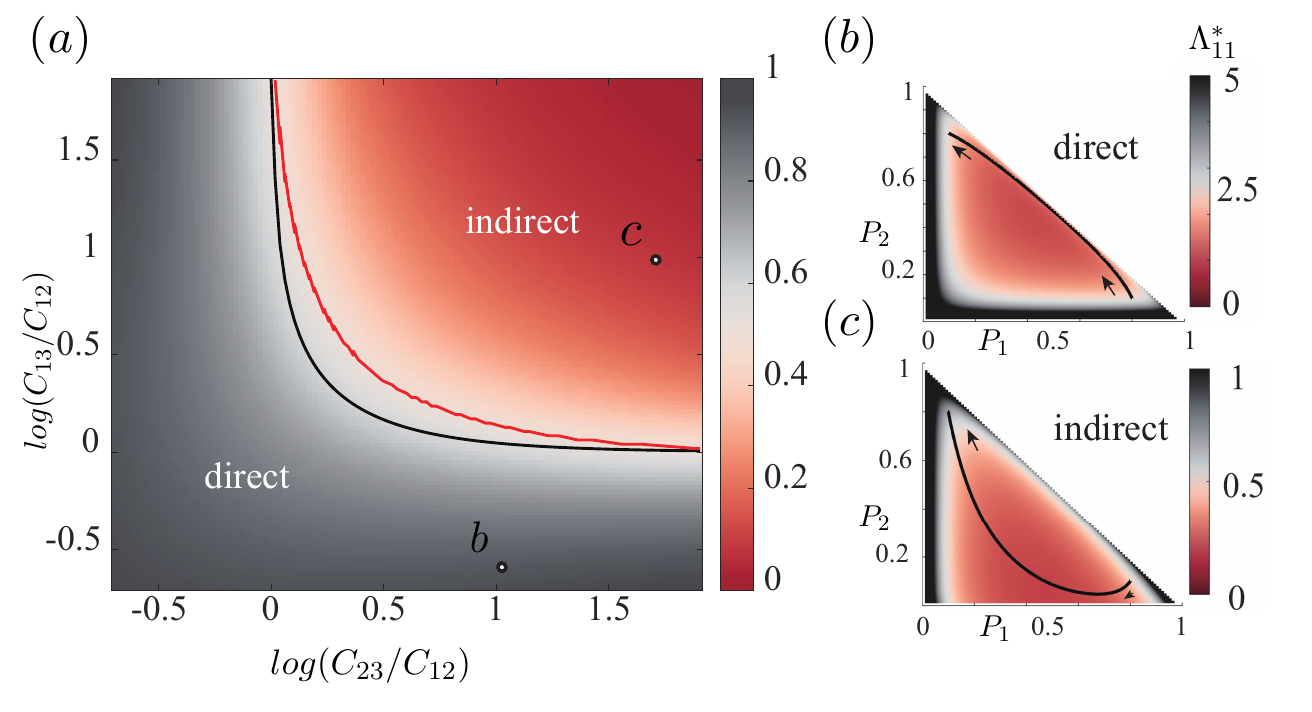}
	\caption{(a) Phase diagram shown for given $C_{12}=5$ and $\tau=5$. The color represents the percentage of transport by the direct link 1-2 versus the indirect route
    1-3-2. The black line is the line where $C_{12}= C_{13}C_{23}/(C_{13}+C_{23})$. The red line is the exact critical line where the percentage is 1/2. (b)\&(c) The reduced metric element $\Lambda^*_{11}(P_1,P_2)$ in the probability space for two sets of parameters ($C_{13}/C_{12}$ and $C_{23}/C_{12}$) marked in (a), which represent the direct and indirect phases, respectively. The geodesics in each case are shown as black lines.  
    }  \label{diagram}
\end{figure}

There are two paths to move from state-1 to state-2: a direct path 1$\rightarrow$2, and an indirect path 1$\rightarrow$3$\rightarrow$2. For simplicity, we fixed the capacity of the direct path $C_{12} = 5$ and studied the optimal transport processes for different $C_{23}$ and $C_{13}$. In Fig.~\ref{diagram}(a), the percentage of direct transport ($1\rightarrow 2$) is shown in the space of relative capacities $(C_{13}/C_{12}, C_{23}/C_{12})$.  
In the lower left corner (black area) where $C_{12} \gg \min(C_{13}, C_{23})$, most of the probability transport goes through the direct channel 1-2; whereas in the upper right corner (red area) where $C_{12} \ll \min(C_{13}, C_{23})$, most of the probability flows through the indirect channel 1-3-2. Interestingly the critical line (red line) where the percentage is $50\%$ is close to the line $C_{12}=C_{13}C_{23}/(C_{13}+C_{23})$ where the effective resistance of the direct and indirect paths are equal.

To understand the modes of optimal probability transport, 
we studied the geodesics in probability space for different choices of the parameters $(C_{13}/C_{12}, C_{23}/C_{12})$. In Fig.~\ref{diagram}(b-c), two typical geodesics are shown in the probability space for the direct and indirect dominated transport regimes, respectively. Generally, the geodesics go through the region where the metric is small to minimize the dissipation length. For the case of direct path dominated transport shown in Fig.~\ref{diagram}(b), the metric is smaller near the diagonal ($P_1+P_2=1$) where $P_3=1-(P_1+P_2)=0$, which attracts the geodesic there leading to a direct path with small $P_3$. But for the indirect transport case shown in Fig.~\ref{diagram}(c), the region of small metric is away from the diagonal, leading to the indirect geodesic path that goes through a region with high $P_3$. See Sec.~I.C and III.A in SM for details of the geodesics in other parts of the parameter space and the full OC ``Phase-Diagram".

\begin{figure}[!ht]
	\centering
	\includegraphics[scale=0.44]{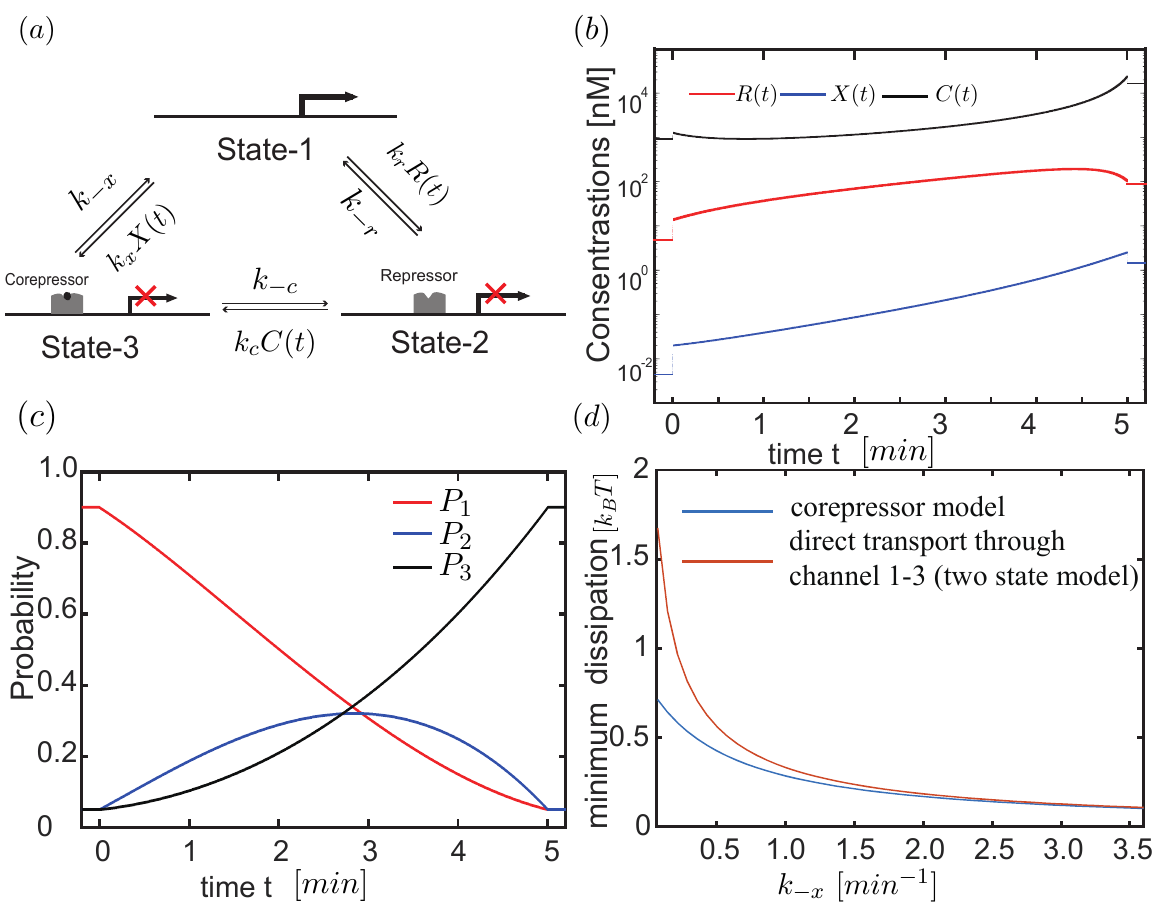}
	\caption{(a) Biochemical network of a repressor-corepressor
model, showing an operator site on DNA in three different states. Transition rate constants are given by \cite{ilker_2022}:$\quad k_{-r}=1.68\mathrm{~min}^{-1}, k_{-c}=0.72\mathrm{~min}^{-1}, k_{-x}=0.072\mathrm{~min}^{-1}$ and $k_r=0.0191 \mathrm{nM}^{-1} \min ^{-1}$, $k_c=$ $7.83 \times 10^{-4} \mathrm{nM}^{-1} \min ^{-1}$, $\quad k_x=0.9 \mathrm{nM}^{-1} \min ^{-1}$; (b) the optimal protocol $X(t)$, $R(t)$, $C(t)$;
(c) the optimal $\textbf{P}(t)$ trajectory; (d) the difference of total entropy production between the original corepressor model and the direct transport model. 
}
\label{corepressor}
\end{figure}

$\textbf{\textit{A 3-state gene circuit}} $. Finally, to show how the $P_i(t)$-dependent $R_{ij}(t)$ affects the optimal protocols in a realistic system, we studied a gene switching model~\cite{ilker_2022}. As shown in Fig.~\ref{corepressor}(a), the gene has three sates, state-1 is the open state that can bind with RNA polymerase to transcribe, state-2 is the partially closed state binding a bare repressor protein, and state-3 is the fully closed state binding a repressor-corepressor complex, which has a much lower dissociation constant compared with that of state-2 ($k_{-x}=0.072\mathrm{~min}^{-1} \ll k_{-r}=1.68\mathrm{~min}^{-1}$).

We consider the switching process from the open state-1 ($P_1=0.9, P_2=0.05, P_3=0.05$) to the fully closed state-3 ($P_1=0.05, P_2=0.05, P_3=0.9$) within a time window $\tau (=5$[min]) after receiving some upstream signals. The transitions between different gene states are mediated by binding/unbinding reactions. 
Here, instead of fixing the total reaction rates $\{C_{ij}\}$ for each binding/unbinding reaction, the dissociation rates $(k_{-r},k_{-c},k_{-x})$ are fixed. The goal of OC is to control the binding rates by controlling the concentrations: $R(t)$ ( bare repressor), $C(t)$ (corepressor), and $X(t)$ (repressor-corepressor complex) in order to switch the gene state with minimum dissipation.

By using the circuit analogy, we find that the resistances for each reaction are: $R_{12}(t)=\frac{1}{2k_{-r}}\frac{1}{P_2(t)}$, $R_{23}(t)=\frac{1}{2k_{-c}}\frac{1}{P_3(t)}$, $R_{13}(t)=\frac{1}{2k_{-x}}\frac{1}{P_3(t)}$ for a given $\textbf{P}(t)$. By solving the OC problem with the 2-step scheme,  
we obtained the optimal protocol and probability trajectory shown in Fig.~\ref{corepressor}(b-c) (see Fig.~S11 in SM for details). The lower dissociation constant in reaction channel 1-3 causes a larger cost for direct transition from state-1 to state-3. Thus, the optimal protocol is to drive most of probability flow through channel 1-2-3 rather than the direct pathway 1-3. 
Although state-2 is not the target state, $P_2$ accumulates transiently as it serves as a buffer to minimize the total dissipation. If we remove state-2, the probability has to transport from state-1 to state-3 directly, and the energy cost for this 2-state model is higher than the 3-state model as shown in Fig.~\ref{corepressor}(d). However, the difference diminishes as $k_{-x}$ increases when the direct channel 1-3 becomes the energetically preferred switching path.

$\textbf{\textit{Discussion}}.$
In this paper, we used a 2-step approach to solve the OC problem in CRNs. In the adiabatic limit, a ``Kirchhoff's law" applies allowing us to show that the total dissipation is determined by a $L_2$-distance (the dissipation length) in the probability space with a dissipation metric that depends on probability and network topology. The optimal probability trajectory and the OC protocol can be obtained by solving the geodesic equation for minimizing this ``dissipation length". Application of our theoretical approach to simple CRNs clearly demonstrates the importance of indirect paths in minimizing the energy cost of changing probability distribution and reveals a rich set of OC protocols depending on details of the network topology and capacity of each reaction. Our approach also leads to a tighter lower bound for the total dissipation that depends on the network topology.

In this paper, we take the adiabatic (slow-driving) limit $\tau\gg \tau_R$, which should be valid in most realistic biological systems as the control time of the whole system is typically slower than the relaxation time $\tau_R$. The minimum control time ($\tau_c$) to achieve the desired change in probability distribution is set by $\tau_R$, i.e., $\tau_c/\tau_R\sim \mathcal{O}(1)$ (see Sec.~III.B in SM for details). 
Our direct numerical results with different values of $\tau$ for the fully-connected 3-node network (Fig.~\ref{fig1}) show that the main conclusions and qualitatively features of the OC protocol and the corresponding optimal transport trajectory found in the adiabatic limit hold true for finite $\tau$ even down to the shortest control time $\tau_c$ (see Sec.~III.B in SM for details). 

The notion of thermodynamic length established $\sim40$ years ago~\cite{Geo_1975,Geo_1983,Geo_1988} has become an important concept in the study of stochastic thermodynamics recently. However, most previous studies focused on the thermodynamic length defined in the Riemannian manifold spanned by (mostly macroscopic) control parameters (e.g., the conjugate forces in the Hamiltonian)~\cite{Geo_2007,Geo_2012,OC_2015,OC_2017,Geo_2020,Geo_2022}, whereas the ``dissipation length" was derived directly in the probability (state) space in our study.

The connection between OC and optimal transport (OT) has been established in systems with continuous state variables governed by Langevin equations where the minimum dissipation is determined by a Wasserstein distance with Euclidean metric in the state space~\cite{OT_2011,OT_2021,OT_2023_2,OT_2023,Dechant_dec1}.  
However, previous attempts in discrete systems have only led to a distance measure that also depends on the control parameters  (transition rates $k_{ij}(t)$)~\cite{Van_2021,Dechant_dec2}. Here,  
by using the Kirchhoff's law, we first determined the optimal rates as a function of probability. A true distance metric tensor $\bf{\Lambda}^*(\textbf{P},\textbf{C})$ in the probability (state) space is then derived, which transforms the OC problem in the rate-space into an OT problem in the state space.  

Our general approach can be applied to study more realistic problems with modifications. In addition to the thermodynamic cost considered in this paper, we can include cost of control itself, which can depend on the rates as well as the rates of changing the rates.  
The number of controllable reactions can also be limited to a small subset of all reactions. An interesting question there is which are the key reactions to control in order to achieve the desired change in probability distribution with minimum cost. Furthermore, the objective function can be extended beyond minimizing entropy production. Other functional costs (or benefits) such as control time $\tau$, the accuracy, and robustness of the control process can also be included in the overall target function. 

$\textbf{\textit{Acknowledgment}}$. Y.T. acknowledges useful discussions with Mark Squillante, David Hathcock, and Qiwei Yu. The work by Y.T. is partially supported by a NIH grant (R35GM137134). The work by Y. Z. and Q. O. is supported by the National Natural Science Foundation of China (Grant No. 12090054) and the Starry Night Science Fund of Zhejiang University Shanghai Institute for Advanced Study. Y.T. would like to thank the Center for Computational Biology at the Flatiron Institute for hospitality while a portion of this work was carried out.
\bibliography{bibliography}

\providecommand{\noopsort}[1]{}\providecommand{\singleletter}[1]{#1}%
\begin{thebibliography}{50}%
\makeatletter
\providecommand \@ifxundefined [1]{%
 \@ifx{#1\undefined}
}%
\providecommand \@ifnum [1]{%
 \ifnum #1\expandafter \@firstoftwo
 \else \expandafter \@secondoftwo
 \fi
}%
\providecommand \@ifx [1]{%
 \ifx #1\expandafter \@firstoftwo
 \else \expandafter \@secondoftwo
 \fi
}%
\providecommand \natexlab [1]{#1}%
\providecommand \enquote  [1]{``#1''}%
\providecommand \bibnamefont  [1]{#1}%
\providecommand \bibfnamefont [1]{#1}%
\providecommand \citenamefont [1]{#1}%
\providecommand \href@noop [0]{\@secondoftwo}%
\providecommand \href [0]{\begingroup \@sanitize@url \@href}%
\providecommand \@href[1]{\@@startlink{#1}\@@href}%
\providecommand \@@href[1]{\endgroup#1\@@endlink}%
\providecommand \@sanitize@url [0]{\catcode `\\12\catcode `\$12\catcode `\&12\catcode `\#12\catcode `\^12\catcode `\_12\catcode `\%12\relax}%
\providecommand \@@startlink[1]{}%
\providecommand \@@endlink[0]{}%
\providecommand \url  [0]{\begingroup\@sanitize@url \@url }%
\providecommand \@url [1]{\endgroup\@href {#1}{\urlprefix }}%
\providecommand \urlprefix  [0]{URL }%
\providecommand \Eprint [0]{\href }%
\providecommand \doibase [0]{https://doi.org/}%
\providecommand \selectlanguage [0]{\@gobble}%
\providecommand \bibinfo  [0]{\@secondoftwo}%
\providecommand \bibfield  [0]{\@secondoftwo}%
\providecommand \translation [1]{[#1]}%
\providecommand \BibitemOpen [0]{}%
\providecommand \bibitemStop [0]{}%
\providecommand \bibitemNoStop [0]{.\EOS\space}%
\providecommand \EOS [0]{\spacefactor3000\relax}%
\providecommand \BibitemShut  [1]{\csname bibitem#1\endcsname}%
\let\auto@bib@innerbib\@empty
\bibitem [{\citenamefont {Sekimoto}(2012)}]{Ken_energetics}%
  \BibitemOpen
  \bibfield  {author} {\bibinfo {author} {\bibfnamefont {K.}~\bibnamefont {Sekimoto}},\ }\href@noop {} {\emph {\bibinfo {title} {Stochastic Energetics}}},\ \bibinfo {edition} {1st}\ ed.\ (\bibinfo  {publisher} {Springer Berlin, Heidelberg},\ \bibinfo {year} {\noopsort{1973c}2012})\BibitemShut {NoStop}%
\bibitem [{\citenamefont {Seifert}(2012)}]{Udo_lambda}%
  \BibitemOpen
  \bibfield  {author} {\bibinfo {author} {\bibfnamefont {U.}~\bibnamefont {Seifert}},\ }\bibfield  {title} {\bibinfo {title} {Stochastic thermodynamics, fluctuation theorems and molecular machines},\ }\bibfield  {journal} {\bibinfo  {journal} {Reports on Progress in Physics}\ }\textbf {\bibinfo {volume} {75}},\ \href {https://doi.org/10.1088/0034-4885/75/12/126001} {10.1088/0034-4885/75/12/126001} (\bibinfo {year} {2012})\BibitemShut {NoStop}%
\bibitem [{\citenamefont {Hopfield}(1974)}]{Hopfield_KPR}%
  \BibitemOpen
  \bibfield  {author} {\bibinfo {author} {\bibfnamefont {J.~J.}\ \bibnamefont {Hopfield}},\ }\bibfield  {title} {\bibinfo {title} {Kinetic proofreading: A new mechanism for reducing errors in biosynthetic processes requiring high specificity},\ }\href {https://doi.org/doi:10.1073/pnas.71.10.4135} {\bibfield  {journal} {\bibinfo  {journal} {Proceedings of the National Academy of Sciences}\ }\textbf {\bibinfo {volume} {71}},\ \bibinfo {pages} {4135} (\bibinfo {year} {1974})}\BibitemShut {NoStop}%
\bibitem [{\citenamefont {Hill}(2012)}]{Hill_transduction}%
  \BibitemOpen
  \bibfield  {author} {\bibinfo {author} {\bibfnamefont {T.~L.}\ \bibnamefont {Hill}},\ }\href@noop {} {\emph {\bibinfo {title} {Free Energy Transduction and Biochemical Cycle Kinetics}}},\ \bibinfo {edition} {1st}\ ed.\ (\bibinfo  {publisher} {Springer New York, NY},\ \bibinfo {year} {\noopsort{1973c}2012})\BibitemShut {NoStop}%
\bibitem [{\citenamefont {Schmiedl}\ and\ \citenamefont {Seifert}(2007{\natexlab{a}})}]{Udo_CRNs}%
  \BibitemOpen
  \bibfield  {author} {\bibinfo {author} {\bibfnamefont {T.}~\bibnamefont {Schmiedl}}\ and\ \bibinfo {author} {\bibfnamefont {U.}~\bibnamefont {Seifert}},\ }\bibfield  {title} {\bibinfo {title} {Stochastic thermodynamics of chemical reaction networks},\ }\href {https://doi.org/10.1063/1.2428297} {\bibfield  {journal} {\bibinfo  {journal} {The Journal of Chemical Physics}\ }\textbf {\bibinfo {volume} {126}},\ \bibinfo {pages} {044101} (\bibinfo {year} {2007}{\natexlab{a}})}\BibitemShut {NoStop}%
\bibitem [{\citenamefont {Rao}\ and\ \citenamefont {Esposito}(2016)}]{Esposito_CRNs}%
  \BibitemOpen
  \bibfield  {author} {\bibinfo {author} {\bibfnamefont {R.}~\bibnamefont {Rao}}\ and\ \bibinfo {author} {\bibfnamefont {M.}~\bibnamefont {Esposito}},\ }\bibfield  {title} {\bibinfo {title} {Nonequilibrium thermodynamics of chemical reaction networks: Wisdom from stochastic thermodynamics},\ }\bibfield  {journal} {\bibinfo  {journal} {Physical Review X}\ }\textbf {\bibinfo {volume} {6}},\ \href {https://doi.org/10.1103/physrevx.6.041064} {10.1103/physrevx.6.041064} (\bibinfo {year} {2016})\BibitemShut {NoStop}%
\bibitem [{\citenamefont {Qian}(2007)}]{Qian_Phosphorylation}%
  \BibitemOpen
  \bibfield  {author} {\bibinfo {author} {\bibfnamefont {H.}~\bibnamefont {Qian}},\ }\bibfield  {title} {\bibinfo {title} {Phosphorylation energy hypothesis: Open chemical systems and their biological functions},\ }\href {https://doi.org/10.1146/annurev.physchem.58.032806.104550} {\bibfield  {journal} {\bibinfo  {journal} {Annual Review of Physical Chemistry}\ }\textbf {\bibinfo {volume} {58}},\ \bibinfo {pages} {113} (\bibinfo {year} {2007})}\BibitemShut {NoStop}%
\bibitem [{\citenamefont {Cao}\ \emph {et~al.}(2015)\citenamefont {Cao}, \citenamefont {Wang}, \citenamefont {Ouyang},\ and\ \citenamefont {Tu}}]{Tu_oscillations}%
  \BibitemOpen
  \bibfield  {author} {\bibinfo {author} {\bibfnamefont {Y.}~\bibnamefont {Cao}}, \bibinfo {author} {\bibfnamefont {H.}~\bibnamefont {Wang}}, \bibinfo {author} {\bibfnamefont {Q.}~\bibnamefont {Ouyang}},\ and\ \bibinfo {author} {\bibfnamefont {Y.}~\bibnamefont {Tu}},\ }\bibfield  {title} {\bibinfo {title} {The free-energy cost of accurate biochemical oscillations},\ }\href {https://doi.org/10.1038/nphys3412} {\bibfield  {journal} {\bibinfo  {journal} {Nature Physics}\ }\textbf {\bibinfo {volume} {11}},\ \bibinfo {pages} {772} (\bibinfo {year} {2015})}\BibitemShut {NoStop}%
\bibitem [{\citenamefont {Lan}\ \emph {et~al.}(2012)\citenamefont {Lan}, \citenamefont {Sartori}, \citenamefont {Neumann}, \citenamefont {Sourjik},\ and\ \citenamefont {Tu}}]{Tu_chemotaxis}%
  \BibitemOpen
  \bibfield  {author} {\bibinfo {author} {\bibfnamefont {G.}~\bibnamefont {Lan}}, \bibinfo {author} {\bibfnamefont {P.}~\bibnamefont {Sartori}}, \bibinfo {author} {\bibfnamefont {S.}~\bibnamefont {Neumann}}, \bibinfo {author} {\bibfnamefont {V.}~\bibnamefont {Sourjik}},\ and\ \bibinfo {author} {\bibfnamefont {Y.}~\bibnamefont {Tu}},\ }\bibfield  {title} {\bibinfo {title} {The energy-speed-accuracy trade-off in sensory adaptation},\ }\href {https://doi.org/10.1038/nphys2276} {\bibfield  {journal} {\bibinfo  {journal} {Nature Physics}\ }\textbf {\bibinfo {volume} {8}},\ \bibinfo {pages} {422} (\bibinfo {year} {2012})}\BibitemShut {NoStop}%
\bibitem [{\citenamefont {Zhang}\ \emph {et~al.}(2020)\citenamefont {Zhang}, \citenamefont {Cao}, \citenamefont {Ouyang},\ and\ \citenamefont {Tu}}]{Tu_synchronization}%
  \BibitemOpen
  \bibfield  {author} {\bibinfo {author} {\bibfnamefont {D.}~\bibnamefont {Zhang}}, \bibinfo {author} {\bibfnamefont {Y.}~\bibnamefont {Cao}}, \bibinfo {author} {\bibfnamefont {Q.}~\bibnamefont {Ouyang}},\ and\ \bibinfo {author} {\bibfnamefont {Y.}~\bibnamefont {Tu}},\ }\bibfield  {title} {\bibinfo {title} {The energy cost and optimal design for synchronization of coupled molecular oscillators},\ }\href {https://doi.org/10.1038/s41567-019-0701-7} {\bibfield  {journal} {\bibinfo  {journal} {Nature Physics}\ }\textbf {\bibinfo {volume} {16}},\ \bibinfo {pages} {95} (\bibinfo {year} {2020})}\BibitemShut {NoStop}%
\bibitem [{\citenamefont {Barato}\ and\ \citenamefont {Seifert}(2015)}]{Udo_TUR}%
  \BibitemOpen
  \bibfield  {author} {\bibinfo {author} {\bibfnamefont {A.~C.}\ \bibnamefont {Barato}}\ and\ \bibinfo {author} {\bibfnamefont {U.}~\bibnamefont {Seifert}},\ }\bibfield  {title} {\bibinfo {title} {Thermodynamic uncertainty relation for biomolecular processes},\ }\bibfield  {journal} {\bibinfo  {journal} {Physical Review Letters}\ }\textbf {\bibinfo {volume} {114}},\ \href {https://doi.org/10.1103/physrevlett.114.158101} {10.1103/physrevlett.114.158101} (\bibinfo {year} {2015})\BibitemShut {NoStop}%
\bibitem [{\citenamefont {Barato}\ and\ \citenamefont {Seifert}(2016)}]{Udo_cost}%
  \BibitemOpen
  \bibfield  {author} {\bibinfo {author} {\bibfnamefont {A.~C.}\ \bibnamefont {Barato}}\ and\ \bibinfo {author} {\bibfnamefont {U.}~\bibnamefont {Seifert}},\ }\bibfield  {title} {\bibinfo {title} {Cost and precision of brownian clocks},\ }\bibfield  {journal} {\bibinfo  {journal} {Physical Review X}\ }\textbf {\bibinfo {volume} {6}},\ \href {https://doi.org/10.1103/physrevx.6.041053} {10.1103/physrevx.6.041053} (\bibinfo {year} {2016})\BibitemShut {NoStop}%
\bibitem [{\citenamefont {Pietzonka}\ and\ \citenamefont {Seifert}(2018)}]{Udo_PEC}%
  \BibitemOpen
  \bibfield  {author} {\bibinfo {author} {\bibfnamefont {P.}~\bibnamefont {Pietzonka}}\ and\ \bibinfo {author} {\bibfnamefont {U.}~\bibnamefont {Seifert}},\ }\bibfield  {title} {\bibinfo {title} {Universal trade-off between power, efficiency, and constancy in steady-state heat engines},\ }\bibfield  {journal} {\bibinfo  {journal} {Physical Review Letters}\ }\textbf {\bibinfo {volume} {120}},\ \href {https://doi.org/10.1103/physrevlett.120.190602} {10.1103/physrevlett.120.190602} (\bibinfo {year} {2018})\BibitemShut {NoStop}%
\bibitem [{\citenamefont {Rodenfels}\ \emph {et~al.}(2019)\citenamefont {Rodenfels}, \citenamefont {Neugebauer},\ and\ \citenamefont {Howard}}]{Howad_oscillations}%
  \BibitemOpen
  \bibfield  {author} {\bibinfo {author} {\bibfnamefont {J.}~\bibnamefont {Rodenfels}}, \bibinfo {author} {\bibfnamefont {K.~M.}\ \bibnamefont {Neugebauer}},\ and\ \bibinfo {author} {\bibfnamefont {J.}~\bibnamefont {Howard}},\ }\bibfield  {title} {\bibinfo {title} {Heat oscillations driven by the embryonic cell cycle reveal the energetic costs of signaling},\ }\href {https://doi.org/10.1016/j.devcel.2018.12.024} {\bibfield  {journal} {\bibinfo  {journal} {Developmental Cell}\ }\textbf {\bibinfo {volume} {48}},\ \bibinfo {pages} {646} (\bibinfo {year} {2019})}\BibitemShut {NoStop}%
\bibitem [{\citenamefont {Del~Vecchio}\ \emph {et~al.}(2016)\citenamefont {Del~Vecchio}, \citenamefont {Dy},\ and\ \citenamefont {Qian}}]{control1}%
  \BibitemOpen
  \bibfield  {author} {\bibinfo {author} {\bibfnamefont {D.}~\bibnamefont {Del~Vecchio}}, \bibinfo {author} {\bibfnamefont {A.~J.}\ \bibnamefont {Dy}},\ and\ \bibinfo {author} {\bibfnamefont {Y.}~\bibnamefont {Qian}},\ }\bibfield  {title} {\bibinfo {title} {Control theory meets synthetic biology},\ }\href {https://doi.org/10.1098/rsif.2016.0380} {\bibfield  {journal} {\bibinfo  {journal} {Journal of The Royal Society Interface}\ }\textbf {\bibinfo {volume} {13}},\ \bibinfo {pages} {20160380} (\bibinfo {year} {2016})}\BibitemShut {NoStop}%
\bibitem [{\citenamefont {Sharp}\ \emph {et~al.}(2021)\citenamefont {Sharp}, \citenamefont {Burrage},\ and\ \citenamefont {Simpson}}]{control2}%
  \BibitemOpen
  \bibfield  {author} {\bibinfo {author} {\bibfnamefont {J.~A.}\ \bibnamefont {Sharp}}, \bibinfo {author} {\bibfnamefont {K.}~\bibnamefont {Burrage}},\ and\ \bibinfo {author} {\bibfnamefont {M.~J.}\ \bibnamefont {Simpson}},\ }\bibfield  {title} {\bibinfo {title} {Implementation and acceleration of optimal control for systems biology},\ }\href {https://doi.org/10.1098/rsif.2021.0241} {\bibfield  {journal} {\bibinfo  {journal} {Journal of The Royal Society Interface}\ }\textbf {\bibinfo {volume} {18}},\ \bibinfo {pages} {20210241} (\bibinfo {year} {2021})}\BibitemShut {NoStop}%
\bibitem [{\citenamefont {Thomas}\ \emph {et~al.}(2019)\citenamefont {Thomas}, \citenamefont {Olufsen}, \citenamefont {Sepulchre}, \citenamefont {Iglesias}, \citenamefont {Ijspeert},\ and\ \citenamefont {Srinivasan}}]{control3}%
  \BibitemOpen
  \bibfield  {author} {\bibinfo {author} {\bibfnamefont {P.~J.}\ \bibnamefont {Thomas}}, \bibinfo {author} {\bibfnamefont {M.}~\bibnamefont {Olufsen}}, \bibinfo {author} {\bibfnamefont {R.}~\bibnamefont {Sepulchre}}, \bibinfo {author} {\bibfnamefont {P.~A.}\ \bibnamefont {Iglesias}}, \bibinfo {author} {\bibfnamefont {A.}~\bibnamefont {Ijspeert}},\ and\ \bibinfo {author} {\bibfnamefont {M.}~\bibnamefont {Srinivasan}},\ }\bibfield  {title} {\bibinfo {title} {Control theory in biology and medicine},\ }\href {https://doi.org/10.1007/s00422-018-00791-5} {\bibfield  {journal} {\bibinfo  {journal} {Biological Cybernetics}\ }\textbf {\bibinfo {volume} {113}},\ \bibinfo {pages} {1} (\bibinfo {year} {2019})}\BibitemShut {NoStop}%
\bibitem [{\citenamefont {Ilker}\ \emph {et~al.}(2022)\citenamefont {Ilker}, \citenamefont {Gungor}, \citenamefont {Kuznets-Speck}, \citenamefont {Chiel}, \citenamefont {Deffner},\ and\ \citenamefont {Hinczewski}}]{ilker_2022}%
  \BibitemOpen
  \bibfield  {author} {\bibinfo {author} {\bibfnamefont {E.}~\bibnamefont {Ilker}}, \bibinfo {author} {\bibfnamefont {O.}~\bibnamefont {Gungor}}, \bibinfo {author} {\bibfnamefont {B.}~\bibnamefont {Kuznets-Speck}}, \bibinfo {author} {\bibfnamefont {J.}~\bibnamefont {Chiel}}, \bibinfo {author} {\bibfnamefont {S.}~\bibnamefont {Deffner}},\ and\ \bibinfo {author} {\bibfnamefont {M.}~\bibnamefont {Hinczewski}},\ }\bibfield  {title} {\bibinfo {title} {Shortcuts in stochastic systems and control of biophysical processes},\ }\bibfield  {journal} {\bibinfo  {journal} {Physical Review X}\ }\textbf {\bibinfo {volume} {12}},\ \href {https://doi.org/10.1103/physrevx.12.021048} {10.1103/physrevx.12.021048} (\bibinfo {year} {2022})\BibitemShut {NoStop}%
\bibitem [{\citenamefont {Schmiedl}\ and\ \citenamefont {Seifert}(2007{\natexlab{b}})}]{OC_2007}%
  \BibitemOpen
  \bibfield  {author} {\bibinfo {author} {\bibfnamefont {T.}~\bibnamefont {Schmiedl}}\ and\ \bibinfo {author} {\bibfnamefont {U.}~\bibnamefont {Seifert}},\ }\bibfield  {title} {\bibinfo {title} {Optimal finite-time processes in stochastic thermodynamics},\ }\bibfield  {journal} {\bibinfo  {journal} {Physical Review Letters}\ }\textbf {\bibinfo {volume} {98}},\ \href {https://doi.org/10.1103/physrevlett.98.108301} {10.1103/physrevlett.98.108301} (\bibinfo {year} {2007}{\natexlab{b}})\BibitemShut {NoStop}%
\bibitem [{\citenamefont {Schmiedl}\ and\ \citenamefont {Seifert}(2007{\natexlab{c}})}]{OC_2008}%
  \BibitemOpen
  \bibfield  {author} {\bibinfo {author} {\bibfnamefont {T.}~\bibnamefont {Schmiedl}}\ and\ \bibinfo {author} {\bibfnamefont {U.}~\bibnamefont {Seifert}},\ }\bibfield  {title} {\bibinfo {title} {Efficiency at maximum power: An analytically solvable model for stochastic heat engines},\ }\href {https://doi.org/10.1209/0295-5075/81/20003} {\bibfield  {journal} {\bibinfo  {journal} {Europhysics Letters}\ }\textbf {\bibinfo {volume} {81}},\ \bibinfo {pages} {20003} (\bibinfo {year} {2007}{\natexlab{c}})}\BibitemShut {NoStop}%
\bibitem [{\citenamefont {Proesmans}\ \emph {et~al.}(2020)\citenamefont {Proesmans}, \citenamefont {Ehrich},\ and\ \citenamefont {Bechhoefer}}]{LP_2020}%
  \BibitemOpen
  \bibfield  {author} {\bibinfo {author} {\bibfnamefont {K.}~\bibnamefont {Proesmans}}, \bibinfo {author} {\bibfnamefont {J.}~\bibnamefont {Ehrich}},\ and\ \bibinfo {author} {\bibfnamefont {J.}~\bibnamefont {Bechhoefer}},\ }\bibfield  {title} {\bibinfo {title} {Finite-time landauer principle},\ }\bibfield  {journal} {\bibinfo  {journal} {Physical Review Letters}\ }\textbf {\bibinfo {volume} {125}},\ \href {https://doi.org/10.1103/physrevlett.125.100602} {10.1103/physrevlett.125.100602} (\bibinfo {year} {2020})\BibitemShut {NoStop}%
\bibitem [{\citenamefont {Rotskoff}\ and\ \citenamefont {Crooks}(2015)}]{OC_2015}%
  \BibitemOpen
  \bibfield  {author} {\bibinfo {author} {\bibfnamefont {G.~M.}\ \bibnamefont {Rotskoff}}\ and\ \bibinfo {author} {\bibfnamefont {G.~E.}\ \bibnamefont {Crooks}},\ }\bibfield  {title} {\bibinfo {title} {Optimal control in nonequilibrium systems: Dynamic riemannian geometry of the ising model},\ }\bibfield  {journal} {\bibinfo  {journal} {Physical Review E}\ }\textbf {\bibinfo {volume} {92}},\ \href {https://doi.org/10.1103/physreve.92.060102} {10.1103/physreve.92.060102} (\bibinfo {year} {2015})\BibitemShut {NoStop}%
\bibitem [{\citenamefont {Rotskoff}\ \emph {et~al.}(2017)\citenamefont {Rotskoff}, \citenamefont {Crooks},\ and\ \citenamefont {Vanden-Eijnden}}]{OC_2017}%
  \BibitemOpen
  \bibfield  {author} {\bibinfo {author} {\bibfnamefont {G.~M.}\ \bibnamefont {Rotskoff}}, \bibinfo {author} {\bibfnamefont {G.~E.}\ \bibnamefont {Crooks}},\ and\ \bibinfo {author} {\bibfnamefont {E.}~\bibnamefont {Vanden-Eijnden}},\ }\bibfield  {title} {\bibinfo {title} {Geometric approach to optimal nonequilibrium control: Minimizing dissipation in nanomagnetic spin systems},\ }\bibfield  {journal} {\bibinfo  {journal} {Physical Review E}\ }\textbf {\bibinfo {volume} {95}},\ \href {https://doi.org/10.1103/physreve.95.012148} {10.1103/physreve.95.012148} (\bibinfo {year} {2017})\BibitemShut {NoStop}%
\bibitem [{\citenamefont {Blaber}\ \emph {et~al.}(2021)\citenamefont {Blaber}, \citenamefont {Louwerse},\ and\ \citenamefont {Sivak}}]{OC_2021}%
  \BibitemOpen
  \bibfield  {author} {\bibinfo {author} {\bibfnamefont {S.}~\bibnamefont {Blaber}}, \bibinfo {author} {\bibfnamefont {M.~D.}\ \bibnamefont {Louwerse}},\ and\ \bibinfo {author} {\bibfnamefont {D.~A.}\ \bibnamefont {Sivak}},\ }\bibfield  {title} {\bibinfo {title} {Steps minimize dissipation in rapidly driven stochastic systems},\ }\bibfield  {journal} {\bibinfo  {journal} {Physical Review E}\ }\textbf {\bibinfo {volume} {104}},\ \href {https://doi.org/10.1103/physreve.104.l022101} {10.1103/physreve.104.l022101} (\bibinfo {year} {2021})\BibitemShut {NoStop}%
\bibitem [{\citenamefont {Remlein}\ and\ \citenamefont {Seifert}(2021)}]{Udo_CRN_2021}%
  \BibitemOpen
  \bibfield  {author} {\bibinfo {author} {\bibfnamefont {B.}~\bibnamefont {Remlein}}\ and\ \bibinfo {author} {\bibfnamefont {U.}~\bibnamefont {Seifert}},\ }\bibfield  {title} {\bibinfo {title} {Optimality of nonconservative driving for finite-time processes with discrete states},\ }\bibfield  {journal} {\bibinfo  {journal} {Physical Review E}\ }\textbf {\bibinfo {volume} {103}},\ \href {https://doi.org/10.1103/physreve.103.l050105} {10.1103/physreve.103.l050105} (\bibinfo {year} {2021})\BibitemShut {NoStop}%
\bibitem [{\citenamefont {Blaber}\ and\ \citenamefont {Sivak}(2022)}]{OC_2022_2}%
  \BibitemOpen
  \bibfield  {author} {\bibinfo {author} {\bibfnamefont {S.}~\bibnamefont {Blaber}}\ and\ \bibinfo {author} {\bibfnamefont {D.~A.}\ \bibnamefont {Sivak}},\ }\bibfield  {title} {\bibinfo {title} {Optimal control with a strong harmonic trap},\ }\bibfield  {journal} {\bibinfo  {journal} {Physical Review E}\ }\textbf {\bibinfo {volume} {106}},\ \href {https://doi.org/10.1103/physreve.106.l022103} {10.1103/physreve.106.l022103} (\bibinfo {year} {2022})\BibitemShut {NoStop}%
\bibitem [{\citenamefont {Blaber}\ and\ \citenamefont {Sivak}(2023)}]{OC_2023}%
  \BibitemOpen
  \bibfield  {author} {\bibinfo {author} {\bibfnamefont {S.}~\bibnamefont {Blaber}}\ and\ \bibinfo {author} {\bibfnamefont {D.~A.}\ \bibnamefont {Sivak}},\ }\bibfield  {title} {\bibinfo {title} {Optimal control in stochastic thermodynamics},\ }\href {https://doi.org/10.1088/2399-6528/acbf04} {\bibfield  {journal} {\bibinfo  {journal} {Journal of Physics Communications}\ }\textbf {\bibinfo {volume} {7}},\ \bibinfo {pages} {033001} (\bibinfo {year} {2023})}\BibitemShut {NoStop}%
\bibitem [{\citenamefont {Aurell}\ \emph {et~al.}(2011)\citenamefont {Aurell}, \citenamefont {Mejía-Monasterio},\ and\ \citenamefont {Muratore-Ginanneschi}}]{OT_2011}%
  \BibitemOpen
  \bibfield  {author} {\bibinfo {author} {\bibfnamefont {E.}~\bibnamefont {Aurell}}, \bibinfo {author} {\bibfnamefont {C.}~\bibnamefont {Mejía-Monasterio}},\ and\ \bibinfo {author} {\bibfnamefont {P.}~\bibnamefont {Muratore-Ginanneschi}},\ }\bibfield  {title} {\bibinfo {title} {Optimal protocols and optimal transport in stochastic thermodynamics},\ }\bibfield  {journal} {\bibinfo  {journal} {Physical Review Letters}\ }\textbf {\bibinfo {volume} {106}},\ \href {https://doi.org/10.1103/physrevlett.106.250601} {10.1103/physrevlett.106.250601} (\bibinfo {year} {2011})\BibitemShut {NoStop}%
\bibitem [{\citenamefont {Nakazato}\ and\ \citenamefont {Ito}(2021)}]{OT_2021}%
  \BibitemOpen
  \bibfield  {author} {\bibinfo {author} {\bibfnamefont {M.}~\bibnamefont {Nakazato}}\ and\ \bibinfo {author} {\bibfnamefont {S.}~\bibnamefont {Ito}},\ }\bibfield  {title} {\bibinfo {title} {Geometrical aspects of entropy production in stochastic thermodynamics based on wasserstein distance},\ }\bibfield  {journal} {\bibinfo  {journal} {Physical Review Research}\ }\textbf {\bibinfo {volume} {3}},\ \href {https://doi.org/10.1103/physrevresearch.3.043093} {10.1103/physrevresearch.3.043093} (\bibinfo {year} {2021})\BibitemShut {NoStop}%
\bibitem [{\citenamefont {Ito}(2023)}]{OT_2023_2}%
  \BibitemOpen
  \bibfield  {author} {\bibinfo {author} {\bibfnamefont {S.}~\bibnamefont {Ito}},\ }\bibfield  {title} {\bibinfo {title} {Geometric thermodynamics for the fokker–planck equation: stochastic thermodynamic links between information geometry and optimal transport},\ }\bibfield  {journal} {\bibinfo  {journal} {Information Geometry}\ }\href {https://doi.org/10.1007/s41884-023-00102-3} {10.1007/s41884-023-00102-3} (\bibinfo {year} {2023})\BibitemShut {NoStop}%
\bibitem [{\citenamefont {Dechant}\ \emph {et~al.}(2022)\citenamefont {Dechant}, \citenamefont {Sasa},\ and\ \citenamefont {Ito}}]{Dechant_dec1}%
  \BibitemOpen
  \bibfield  {author} {\bibinfo {author} {\bibfnamefont {A.}~\bibnamefont {Dechant}}, \bibinfo {author} {\bibfnamefont {S.-I.}\ \bibnamefont {Sasa}},\ and\ \bibinfo {author} {\bibfnamefont {S.}~\bibnamefont {Ito}},\ }\bibfield  {title} {\bibinfo {title} {Geometric decomposition of entropy production in out-of-equilibrium systems},\ }\bibfield  {journal} {\bibinfo  {journal} {Physical Review Research}\ }\textbf {\bibinfo {volume} {4}},\ \href {https://doi.org/10.1103/physrevresearch.4.l012034} {10.1103/physrevresearch.4.l012034} (\bibinfo {year} {2022})\BibitemShut {NoStop}%
\bibitem [{\citenamefont {Weinhold}(1975)}]{Geo_1975}%
  \BibitemOpen
  \bibfield  {author} {\bibinfo {author} {\bibfnamefont {F.}~\bibnamefont {Weinhold}},\ }\bibfield  {title} {\bibinfo {title} {Metric geometry of equilibrium thermodynamics},\ }\href {https://doi.org/10.1063/1.431689} {\bibfield  {journal} {\bibinfo  {journal} {The Journal of Chemical Physics}\ }\textbf {\bibinfo {volume} {63}},\ \bibinfo {pages} {2479} (\bibinfo {year} {1975})}\BibitemShut {NoStop}%
\bibitem [{\citenamefont {Salamon}\ and\ \citenamefont {Berry}(1983)}]{Geo_1983}%
  \BibitemOpen
  \bibfield  {author} {\bibinfo {author} {\bibfnamefont {P.}~\bibnamefont {Salamon}}\ and\ \bibinfo {author} {\bibfnamefont {R.~S.}\ \bibnamefont {Berry}},\ }\bibfield  {title} {\bibinfo {title} {Thermodynamic length and dissipated availability},\ }\href {https://doi.org/10.1103/physrevlett.51.1127} {\bibfield  {journal} {\bibinfo  {journal} {Physical Review Letters}\ }\textbf {\bibinfo {volume} {51}},\ \bibinfo {pages} {1127} (\bibinfo {year} {1983})}\BibitemShut {NoStop}%
\bibitem [{\citenamefont {Andresen}\ \emph {et~al.}(1988)\citenamefont {Andresen}, \citenamefont {Berry}, \citenamefont {Gilmore}, \citenamefont {Ihrig},\ and\ \citenamefont {Salamon}}]{Geo_1988}%
  \BibitemOpen
  \bibfield  {author} {\bibinfo {author} {\bibfnamefont {B.}~\bibnamefont {Andresen}}, \bibinfo {author} {\bibfnamefont {R.~S.}\ \bibnamefont {Berry}}, \bibinfo {author} {\bibfnamefont {R.}~\bibnamefont {Gilmore}}, \bibinfo {author} {\bibfnamefont {E.}~\bibnamefont {Ihrig}},\ and\ \bibinfo {author} {\bibfnamefont {P.}~\bibnamefont {Salamon}},\ }\bibfield  {title} {\bibinfo {title} {Thermodynamic geometry and the metrics of weinhold and gilmore},\ }\href {https://doi.org/10.1103/physreva.37.845} {\bibfield  {journal} {\bibinfo  {journal} {Physical Review A}\ }\textbf {\bibinfo {volume} {37}},\ \bibinfo {pages} {845} (\bibinfo {year} {1988})}\BibitemShut {NoStop}%
\bibitem [{\citenamefont {Crooks}(2007)}]{Geo_2007}%
  \BibitemOpen
  \bibfield  {author} {\bibinfo {author} {\bibfnamefont {G.~E.}\ \bibnamefont {Crooks}},\ }\bibfield  {title} {\bibinfo {title} {Measuring thermodynamic length},\ }\bibfield  {journal} {\bibinfo  {journal} {Physical Review Letters}\ }\textbf {\bibinfo {volume} {99}},\ \href {https://doi.org/10.1103/physrevlett.99.100602} {10.1103/physrevlett.99.100602} (\bibinfo {year} {2007})\BibitemShut {NoStop}%
\bibitem [{\citenamefont {Sivak}\ and\ \citenamefont {Crooks}(2012)}]{Geo_2012}%
  \BibitemOpen
  \bibfield  {author} {\bibinfo {author} {\bibfnamefont {D.~A.}\ \bibnamefont {Sivak}}\ and\ \bibinfo {author} {\bibfnamefont {G.~E.}\ \bibnamefont {Crooks}},\ }\bibfield  {title} {\bibinfo {title} {Thermodynamic metrics and optimal paths},\ }\bibfield  {journal} {\bibinfo  {journal} {Physical Review Letters}\ }\textbf {\bibinfo {volume} {108}},\ \href {https://doi.org/10.1103/physrevlett.108.190602} {10.1103/physrevlett.108.190602} (\bibinfo {year} {2012})\BibitemShut {NoStop}%
\bibitem [{\citenamefont {Ma}\ \emph {et~al.}(2020)\citenamefont {Ma}, \citenamefont {Zhai}, \citenamefont {Chen}, \citenamefont {Sun},\ and\ \citenamefont {Dong}}]{Geo_2020}%
  \BibitemOpen
  \bibfield  {author} {\bibinfo {author} {\bibfnamefont {Y.-H.}\ \bibnamefont {Ma}}, \bibinfo {author} {\bibfnamefont {R.-X.}\ \bibnamefont {Zhai}}, \bibinfo {author} {\bibfnamefont {J.}~\bibnamefont {Chen}}, \bibinfo {author} {\bibfnamefont {C.~P.}\ \bibnamefont {Sun}},\ and\ \bibinfo {author} {\bibfnamefont {H.}~\bibnamefont {Dong}},\ }\bibfield  {title} {\bibinfo {title} {Experimental test of the 1/$\tau$-scaling entropy generation in finite-time thermodynamics},\ }\bibfield  {journal} {\bibinfo  {journal} {Physical Review Letters}\ }\textbf {\bibinfo {volume} {125}},\ \href {https://doi.org/10.1103/physrevlett.125.210601} {10.1103/physrevlett.125.210601} (\bibinfo {year} {2020})\BibitemShut {NoStop}%
\bibitem [{\citenamefont {Li}\ \emph {et~al.}(2022)\citenamefont {Li}, \citenamefont {Chen}, \citenamefont {Sun},\ and\ \citenamefont {Dong}}]{Geo_2022}%
  \BibitemOpen
  \bibfield  {author} {\bibinfo {author} {\bibfnamefont {G.}~\bibnamefont {Li}}, \bibinfo {author} {\bibfnamefont {J.-F.}\ \bibnamefont {Chen}}, \bibinfo {author} {\bibfnamefont {C.~P.}\ \bibnamefont {Sun}},\ and\ \bibinfo {author} {\bibfnamefont {H.}~\bibnamefont {Dong}},\ }\bibfield  {title} {\bibinfo {title} {Geodesic path for the minimal energy cost in shortcuts to isothermality},\ }\bibfield  {journal} {\bibinfo  {journal} {Physical Review Letters}\ }\textbf {\bibinfo {volume} {128}},\ \href {https://doi.org/10.1103/physrevlett.128.230603} {10.1103/physrevlett.128.230603} (\bibinfo {year} {2022})\BibitemShut {NoStop}%
\bibitem [{\citenamefont {Shiraishi}\ \emph {et~al.}(2018)\citenamefont {Shiraishi}, \citenamefont {Funo},\ and\ \citenamefont {Saito}}]{SL_2018}%
  \BibitemOpen
  \bibfield  {author} {\bibinfo {author} {\bibfnamefont {N.}~\bibnamefont {Shiraishi}}, \bibinfo {author} {\bibfnamefont {K.}~\bibnamefont {Funo}},\ and\ \bibinfo {author} {\bibfnamefont {K.}~\bibnamefont {Saito}},\ }\bibfield  {title} {\bibinfo {title} {Speed limit for classical stochastic processes},\ }\bibfield  {journal} {\bibinfo  {journal} {Physical Review Letters}\ }\textbf {\bibinfo {volume} {121}},\ \href {https://doi.org/10.1103/physrevlett.121.070601} {10.1103/physrevlett.121.070601} (\bibinfo {year} {2018})\BibitemShut {NoStop}%
\bibitem [{\citenamefont {Van~Vu}\ and\ \citenamefont {Hasegawa}(2021)}]{Van_2021}%
  \BibitemOpen
  \bibfield  {author} {\bibinfo {author} {\bibfnamefont {T.}~\bibnamefont {Van~Vu}}\ and\ \bibinfo {author} {\bibfnamefont {Y.}~\bibnamefont {Hasegawa}},\ }\bibfield  {title} {\bibinfo {title} {Geometrical bounds of the irreversibility in markovian systems},\ }\bibfield  {journal} {\bibinfo  {journal} {Physical Review Letters}\ }\textbf {\bibinfo {volume} {126}},\ \href {https://doi.org/10.1103/physrevlett.126.010601} {10.1103/physrevlett.126.010601} (\bibinfo {year} {2021})\BibitemShut {NoStop}%
\bibitem [{\citenamefont {Yoshimura}\ and\ \citenamefont {Ito}(2021)}]{CRNs_2021}%
  \BibitemOpen
  \bibfield  {author} {\bibinfo {author} {\bibfnamefont {K.}~\bibnamefont {Yoshimura}}\ and\ \bibinfo {author} {\bibfnamefont {S.}~\bibnamefont {Ito}},\ }\bibfield  {title} {\bibinfo {title} {Thermodynamic uncertainty relation and thermodynamic speed limit in deterministic chemical reaction networks},\ }\bibfield  {journal} {\bibinfo  {journal} {Physical Review Letters}\ }\textbf {\bibinfo {volume} {127}},\ \href {https://doi.org/10.1103/physrevlett.127.160601} {10.1103/physrevlett.127.160601} (\bibinfo {year} {2021})\BibitemShut {NoStop}%
\bibitem [{\citenamefont {Dechant}(2022)}]{OT_2022}%
  \BibitemOpen
  \bibfield  {author} {\bibinfo {author} {\bibfnamefont {A.}~\bibnamefont {Dechant}},\ }\bibfield  {title} {\bibinfo {title} {Minimum entropy production, detailed balance and wasserstein distance for continuous-time markov processes},\ }\href {https://doi.org/10.1088/1751-8121/ac4ac0} {\bibfield  {journal} {\bibinfo  {journal} {Journal of Physics A: Mathematical and Theoretical}\ }\textbf {\bibinfo {volume} {55}},\ \bibinfo {pages} {094001} (\bibinfo {year} {2022})}\BibitemShut {NoStop}%
\bibitem [{\citenamefont {Van~Vu}\ and\ \citenamefont {Saito}(2023)}]{OT_2023}%
  \BibitemOpen
  \bibfield  {author} {\bibinfo {author} {\bibfnamefont {T.}~\bibnamefont {Van~Vu}}\ and\ \bibinfo {author} {\bibfnamefont {K.}~\bibnamefont {Saito}},\ }\bibfield  {title} {\bibinfo {title} {Thermodynamic unification of optimal transport: Thermodynamic uncertainty relation, minimum dissipation, and thermodynamic speed limits},\ }\bibfield  {journal} {\bibinfo  {journal} {Physical Review X}\ }\textbf {\bibinfo {volume} {13}},\ \href {https://doi.org/10.1103/physrevx.13.011013} {10.1103/physrevx.13.011013} (\bibinfo {year} {2023})\BibitemShut {NoStop}%
\bibitem [{\citenamefont {Yoshimura}\ \emph {et~al.}(2023)\citenamefont {Yoshimura}, \citenamefont {Kolchinsky}, \citenamefont {Dechant},\ and\ \citenamefont {Ito}}]{Dechant_dec2}%
  \BibitemOpen
  \bibfield  {author} {\bibinfo {author} {\bibfnamefont {K.}~\bibnamefont {Yoshimura}}, \bibinfo {author} {\bibfnamefont {A.}~\bibnamefont {Kolchinsky}}, \bibinfo {author} {\bibfnamefont {A.}~\bibnamefont {Dechant}},\ and\ \bibinfo {author} {\bibfnamefont {S.}~\bibnamefont {Ito}},\ }\bibfield  {title} {\bibinfo {title} {Housekeeping and excess entropy production for general nonlinear dynamics},\ }\bibfield  {journal} {\bibinfo  {journal} {Physical Review Research}\ }\textbf {\bibinfo {volume} {5}},\ \href {https://doi.org/10.1103/physrevresearch.5.013017} {10.1103/physrevresearch.5.013017} (\bibinfo {year} {2023})\BibitemShut {NoStop}%
\bibitem [{\citenamefont {Muratore-Ginanneschi}\ \emph {et~al.}(2013)\citenamefont {Muratore-Ginanneschi}, \citenamefont {Mejía-Monasterio},\ and\ \citenamefont {Peliti}}]{DO_2013}%
  \BibitemOpen
  \bibfield  {author} {\bibinfo {author} {\bibfnamefont {P.}~\bibnamefont {Muratore-Ginanneschi}}, \bibinfo {author} {\bibfnamefont {C.}~\bibnamefont {Mejía-Monasterio}},\ and\ \bibinfo {author} {\bibfnamefont {L.}~\bibnamefont {Peliti}},\ }\bibfield  {title} {\bibinfo {title} {Heat release by controlled continuous-time markov jump processes},\ }\href {https://doi.org/10.1007/s10955-012-0676-6} {\bibfield  {journal} {\bibinfo  {journal} {Journal of Statistical Physics}\ }\textbf {\bibinfo {volume} {150}},\ \bibinfo {pages} {181} (\bibinfo {year} {2013})}\BibitemShut {NoStop}%
\bibitem [{Note1()}]{Note1}%
  \BibitemOpen
  \bibinfo {note} {In the adiabatic limit, the control process is carried out near-equilibrium: $P_ik_{ij}\approx P_jk_{ji} (1+\protect \mathcal {O}(\tau _R/\tau ))$. Together with the constraint $k_{ij}+k_{ji}=C_{ij}$, we have $R_{ij}\approx \protect \frac {1}{2C_{ij}}(\protect \frac {1}{P_i(t)}+\protect \frac {1}{P_j(t)})$ to the leading order in $\protect \mathcal {O}(\tau _R/\tau )$.}\BibitemShut {Stop}%
\bibitem [{\citenamefont {Lin}(2020)}]{Circuit_2020}%
  \BibitemOpen
  \bibfield  {author} {\bibinfo {author} {\bibfnamefont {M.~M.}\ \bibnamefont {Lin}},\ }\bibfield  {title} {\bibinfo {title} {Circuit reduction of heterogeneous nonequilibrium systems},\ }\bibfield  {journal} {\bibinfo  {journal} {Physical Review Letters}\ }\textbf {\bibinfo {volume} {125}},\ \href {https://doi.org/10.1103/physrevlett.125.218101} {10.1103/physrevlett.125.218101} (\bibinfo {year} {2020})\BibitemShut {NoStop}%
\bibitem [{\citenamefont {Avanzini}\ \emph {et~al.}(2023)\citenamefont {Avanzini}, \citenamefont {Freitas},\ and\ \citenamefont {Esposito}}]{Circuit_2023}%
  \BibitemOpen
  \bibfield  {author} {\bibinfo {author} {\bibfnamefont {F.}~\bibnamefont {Avanzini}}, \bibinfo {author} {\bibfnamefont {N.}~\bibnamefont {Freitas}},\ and\ \bibinfo {author} {\bibfnamefont {M.}~\bibnamefont {Esposito}},\ }\bibfield  {title} {\bibinfo {title} {Circuit theory for chemical reaction networks},\ }\bibfield  {journal} {\bibinfo  {journal} {Physical Review X}\ }\textbf {\bibinfo {volume} {13}},\ \href {https://doi.org/10.1103/physrevx.13.021041} {10.1103/physrevx.13.021041} (\bibinfo {year} {2023})\BibitemShut {NoStop}%
\bibitem [{Note2()}]{Note2}%
  \BibitemOpen
  \bibinfo {note} {The inverse here is formal, since $\protect \bf {\Gamma }(\protect \textbf {R})$ is a singular matrix due to $\DOTSB \sum@ \slimits@ _iJ_i(t)=\DOTSB \sum@ \slimits@ _i\protect \dot {P}_i(t)=0$, see Sec.~I.A in SM for the definition of the inverse}\BibitemShut {NoStop}%
\bibitem [{Note3()}]{Note3}%
  \BibitemOpen
  \bibinfo {note} {In the expression for $k_{ij}^*=(P_i+P_j)^{-1} (P_j C_{ij}+\DOTSB \sum@ \slimits@ _{k}\Phi _{ij,k}\protect \dot {P}_k)$, the first term $(P_i+P_j)^{-1} P_j C_{ij}$ represents the equilibrium effect when detailed balance is satisfied, and the second term $(P_i+P_j)^{-1} \DOTSB \sum@ \slimits@ _{k}\Phi _{ij,k}\protect \dot {P}_k$, which is of order $\protect \mathcal {O}(\tau _R/\tau )$, represents the out-of-equilibrium effect that breaks detailed balance.}\BibitemShut {Stop}%
\end{thebibliography}%


\providecommand{\noopsort}[1]{}\providecommand{\singleletter}[1]{#1}%
\begin{thebibliography}{12}%
\makeatletter
\providecommand \@ifxundefined [1]{%
 \@ifx{#1\undefined}
}%
\providecommand \@ifnum [1]{%
 \ifnum #1\expandafter \@firstoftwo
 \else \expandafter \@secondoftwo
 \fi
}%
\providecommand \@ifx [1]{%
 \ifx #1\expandafter \@firstoftwo
 \else \expandafter \@secondoftwo
 \fi
}%
\providecommand \natexlab [1]{#1}%
\providecommand \enquote  [1]{``#1''}%
\providecommand \bibnamefont  [1]{#1}%
\providecommand \bibfnamefont [1]{#1}%
\providecommand \citenamefont [1]{#1}%
\providecommand \href@noop [0]{\@secondoftwo}%
\providecommand \href [0]{\begingroup \@sanitize@url \@href}%
\providecommand \@href[1]{\@@startlink{#1}\@@href}%
\providecommand \@@href[1]{\endgroup#1\@@endlink}%
\providecommand \@sanitize@url [0]{\catcode `\\12\catcode `\$12\catcode `\&12\catcode `\#12\catcode `\^12\catcode `\_12\catcode `\%12\relax}%
\providecommand \@@startlink[1]{}%
\providecommand \@@endlink[0]{}%
\providecommand \url  [0]{\begingroup\@sanitize@url \@url }%
\providecommand \@url [1]{\endgroup\@href {#1}{\urlprefix }}%
\providecommand \urlprefix  [0]{URL }%
\providecommand \Eprint [0]{\href }%
\providecommand \doibase [0]{https://doi.org/}%
\providecommand \selectlanguage [0]{\@gobble}%
\providecommand \bibinfo  [0]{\@secondoftwo}%
\providecommand \bibfield  [0]{\@secondoftwo}%
\providecommand \translation [1]{[#1]}%
\providecommand \BibitemOpen [0]{}%
\providecommand \bibitemStop [0]{}%
\providecommand \bibitemNoStop [0]{.\EOS\space}%
\providecommand \EOS [0]{\spacefactor3000\relax}%
\providecommand \BibitemShut  [1]{\csname bibitem#1\endcsname}%
\let\auto@bib@innerbib\@empty
\bibitem [{\citenamefont {Ge}\ and\ \citenamefont {Qian}(2010)}]{qian_2010}%
  \BibitemOpen
  \bibfield  {author} {\bibinfo {author} {\bibfnamefont {H.}~\bibnamefont {Ge}}\ and\ \bibinfo {author} {\bibfnamefont {H.}~\bibnamefont {Qian}},\ }\bibfield  {title} {\bibinfo {title} {Physical origins of entropy production, free energy dissipation, and their mathematical representations},\ }\bibfield  {journal} {\bibinfo  {journal} {Physical Review E}\ }\textbf {\bibinfo {volume} {81}},\ \href {https://doi.org/10.1103/physreve.81.051133} {10.1103/physreve.81.051133} (\bibinfo {year} {2010})\BibitemShut {NoStop}%
\bibitem [{\citenamefont {Salamon}\ and\ \citenamefont {Berry}(1983)}]{Geo_1983}%
  \BibitemOpen
  \bibfield  {author} {\bibinfo {author} {\bibfnamefont {P.}~\bibnamefont {Salamon}}\ and\ \bibinfo {author} {\bibfnamefont {R.~S.}\ \bibnamefont {Berry}},\ }\bibfield  {title} {\bibinfo {title} {Thermodynamic length and dissipated availability},\ }\href {https://doi.org/10.1103/physrevlett.51.1127} {\bibfield  {journal} {\bibinfo  {journal} {Physical Review Letters}\ }\textbf {\bibinfo {volume} {51}},\ \bibinfo {pages} {1127} (\bibinfo {year} {1983})}\BibitemShut {NoStop}%
\bibitem [{\citenamefont {Crooks}(2007)}]{Geo_2007}%
  \BibitemOpen
  \bibfield  {author} {\bibinfo {author} {\bibfnamefont {G.~E.}\ \bibnamefont {Crooks}},\ }\bibfield  {title} {\bibinfo {title} {Measuring thermodynamic length},\ }\bibfield  {journal} {\bibinfo  {journal} {Physical Review Letters}\ }\textbf {\bibinfo {volume} {99}},\ \href {https://doi.org/10.1103/physrevlett.99.100602} {10.1103/physrevlett.99.100602} (\bibinfo {year} {2007})\BibitemShut {NoStop}%
\bibitem [{\citenamefont {Sivak}\ and\ \citenamefont {Crooks}(2012)}]{Geo_2012}%
  \BibitemOpen
  \bibfield  {author} {\bibinfo {author} {\bibfnamefont {D.~A.}\ \bibnamefont {Sivak}}\ and\ \bibinfo {author} {\bibfnamefont {G.~E.}\ \bibnamefont {Crooks}},\ }\bibfield  {title} {\bibinfo {title} {Thermodynamic metrics and optimal paths},\ }\bibfield  {journal} {\bibinfo  {journal} {Physical Review Letters}\ }\textbf {\bibinfo {volume} {108}},\ \href {https://doi.org/10.1103/physrevlett.108.190602} {10.1103/physrevlett.108.190602} (\bibinfo {year} {2012})\BibitemShut {NoStop}%
\bibitem [{\citenamefont {Rotskoff}\ and\ \citenamefont {Crooks}(2015)}]{OC_2015}%
  \BibitemOpen
  \bibfield  {author} {\bibinfo {author} {\bibfnamefont {G.~M.}\ \bibnamefont {Rotskoff}}\ and\ \bibinfo {author} {\bibfnamefont {G.~E.}\ \bibnamefont {Crooks}},\ }\bibfield  {title} {\bibinfo {title} {Optimal control in nonequilibrium systems: Dynamic riemannian geometry of the ising model},\ }\bibfield  {journal} {\bibinfo  {journal} {Physical Review E}\ }\textbf {\bibinfo {volume} {92}},\ \href {https://doi.org/10.1103/physreve.92.060102} {10.1103/physreve.92.060102} (\bibinfo {year} {2015})\BibitemShut {NoStop}%
\bibitem [{\citenamefont {Rotskoff}\ \emph {et~al.}(2017)\citenamefont {Rotskoff}, \citenamefont {Crooks},\ and\ \citenamefont {Vanden-Eijnden}}]{OC_2017}%
  \BibitemOpen
  \bibfield  {author} {\bibinfo {author} {\bibfnamefont {G.~M.}\ \bibnamefont {Rotskoff}}, \bibinfo {author} {\bibfnamefont {G.~E.}\ \bibnamefont {Crooks}},\ and\ \bibinfo {author} {\bibfnamefont {E.}~\bibnamefont {Vanden-Eijnden}},\ }\bibfield  {title} {\bibinfo {title} {Geometric approach to optimal nonequilibrium control: Minimizing dissipation in nanomagnetic spin systems},\ }\bibfield  {journal} {\bibinfo  {journal} {Physical Review E}\ }\textbf {\bibinfo {volume} {95}},\ \href {https://doi.org/10.1103/physreve.95.012148} {10.1103/physreve.95.012148} (\bibinfo {year} {2017})\BibitemShut {NoStop}%
\bibitem [{\citenamefont {Shiraishi}\ \emph {et~al.}(2018)\citenamefont {Shiraishi}, \citenamefont {Funo},\ and\ \citenamefont {Saito}}]{SL_2018}%
  \BibitemOpen
  \bibfield  {author} {\bibinfo {author} {\bibfnamefont {N.}~\bibnamefont {Shiraishi}}, \bibinfo {author} {\bibfnamefont {K.}~\bibnamefont {Funo}},\ and\ \bibinfo {author} {\bibfnamefont {K.}~\bibnamefont {Saito}},\ }\bibfield  {title} {\bibinfo {title} {Speed limit for classical stochastic processes},\ }\bibfield  {journal} {\bibinfo  {journal} {Physical Review Letters}\ }\textbf {\bibinfo {volume} {121}},\ \href {https://doi.org/10.1103/physrevlett.121.070601} {10.1103/physrevlett.121.070601} (\bibinfo {year} {2018})\BibitemShut {NoStop}%
\bibitem [{\citenamefont {Dechant}(2022)}]{OT_2022}%
  \BibitemOpen
  \bibfield  {author} {\bibinfo {author} {\bibfnamefont {A.}~\bibnamefont {Dechant}},\ }\bibfield  {title} {\bibinfo {title} {Minimum entropy production, detailed balance and wasserstein distance for continuous-time markov processes},\ }\href {https://doi.org/10.1088/1751-8121/ac4ac0} {\bibfield  {journal} {\bibinfo  {journal} {Journal of Physics A: Mathematical and Theoretical}\ }\textbf {\bibinfo {volume} {55}},\ \bibinfo {pages} {094001} (\bibinfo {year} {2022})}\BibitemShut {NoStop}%
\bibitem [{\citenamefont {Van~Vu}\ and\ \citenamefont {Saito}(2023)}]{OT_2023}%
  \BibitemOpen
  \bibfield  {author} {\bibinfo {author} {\bibfnamefont {T.}~\bibnamefont {Van~Vu}}\ and\ \bibinfo {author} {\bibfnamefont {K.}~\bibnamefont {Saito}},\ }\bibfield  {title} {\bibinfo {title} {Thermodynamic unification of optimal transport: Thermodynamic uncertainty relation, minimum dissipation, and thermodynamic speed limits},\ }\bibfield  {journal} {\bibinfo  {journal} {Physical Review X}\ }\textbf {\bibinfo {volume} {13}},\ \href {https://doi.org/10.1103/physrevx.13.011013} {10.1103/physrevx.13.011013} (\bibinfo {year} {2023})\BibitemShut {NoStop}%
\bibitem [{\citenamefont {Schmiedl}\ and\ \citenamefont {Seifert}(2007)}]{OC_2007}%
  \BibitemOpen
  \bibfield  {author} {\bibinfo {author} {\bibfnamefont {T.}~\bibnamefont {Schmiedl}}\ and\ \bibinfo {author} {\bibfnamefont {U.}~\bibnamefont {Seifert}},\ }\bibfield  {title} {\bibinfo {title} {Optimal finite-time processes in stochastic thermodynamics},\ }\bibfield  {journal} {\bibinfo  {journal} {Physical Review Letters}\ }\textbf {\bibinfo {volume} {98}},\ \href {https://doi.org/10.1103/physrevlett.98.108301} {10.1103/physrevlett.98.108301} (\bibinfo {year} {2007})\BibitemShut {NoStop}%
\bibitem [{\citenamefont {Blaber}\ \emph {et~al.}(2021)\citenamefont {Blaber}, \citenamefont {Louwerse},\ and\ \citenamefont {Sivak}}]{OC_2021}%
  \BibitemOpen
  \bibfield  {author} {\bibinfo {author} {\bibfnamefont {S.}~\bibnamefont {Blaber}}, \bibinfo {author} {\bibfnamefont {M.~D.}\ \bibnamefont {Louwerse}},\ and\ \bibinfo {author} {\bibfnamefont {D.~A.}\ \bibnamefont {Sivak}},\ }\bibfield  {title} {\bibinfo {title} {Steps minimize dissipation in rapidly driven stochastic systems},\ }\bibfield  {journal} {\bibinfo  {journal} {Physical Review E}\ }\textbf {\bibinfo {volume} {104}},\ \href {https://doi.org/10.1103/physreve.104.l022101} {10.1103/physreve.104.l022101} (\bibinfo {year} {2021})\BibitemShut {NoStop}%
\bibitem [{Note1()}]{Note1}%
  \BibitemOpen
  \bibinfo {note} {Http://www.ee.ic.ac.uk/ICLOCS/default.htm}\BibitemShut {NoStop}%
\end{thebibliography}%
\end{document}


\title{Supplementary Materials}

\author{Yikuan Zhang}
\affiliation{%
School of Physics, Peking University, Beijing 100871, China 
}%
\author{Qi Ouyang}%
\affiliation{%
Institute for Advanced Study in Physics, Zhejiang University, Hangzhou 310058, China
}%
\affiliation{%
Center for Quantitative Biology,
AAIC, Peking University, Beijing 100871, China
}%
\author{Yuhai Tu}%
\affiliation{IBM T. J. Watson Research Center,
Yorktown Heights, New York 10598, USA}

\date{\today}

\maketitle
\tableofcontents

\newpage
\section{The geometry of optimal control for discrete state variables}
\subsection{The inverse of the $\bf{\Gamma}$ matrix}
Since $\dot{\textbf{P}}=\bf{\Gamma}\textbf{V}$ and $\sum_iJ_i(t)=\sum_i\dot{P}_i(t)=0$, $\bf{\Gamma}(\textbf{R})$ is a singular matrix with $rank(\bf{\Gamma})=dim(\bf{\Gamma})-1$. The way to get $\bf{\Omega}\equiv\bf{\Gamma}^{-1}$ is to reduce the dimension.
Since the potential $V_{i}$ for each node-i plus or minus an arbitrary constant simultaneously will not change the result of currents, we can set $V_{k}=0$ for an arbitrary node k. For simplicity, here we chose $k=N$. At the same time, wiping out the N-th row and column of matrix $\bf{\Gamma}$, we get the resulting invertible matrix $\bf{\gamma}$. In principle, there are infinite ways to reconstruct the matrix $\bf{\Omega}$ through augmenting $\gamma^{-1}$ into a $(N\times N)$ matrix as long as it satisfies $\Omega_{iN}=0$ and $\Omega_{Ni}=\Omega_{NN}$ for $i\neq N$. Different ways reconstructing $\bf{\Omega}$ lead to different expressions of $\bf{\Lambda}$. Here, we chose to set all the elements to 0 except $\gamma^{-1}$:

\begin{equation}
    \bf{\Omega}=\bf{\Gamma}^{-1} \equiv \begin{bmatrix} \bf{\gamma}^{-1} & 0 \\ 0 & 0 \end{bmatrix},
\end{equation}
leading to the elements in the N-th column of $\bf{\Phi}$ are 0 (i.e. $\Phi_{ij,N}=0$ for any $ij$). As a consequence, the elements in the N-th row and column of $\bf{\Lambda}$ are 0 due to $\Lambda_{ij}(\textbf{R})= \Sigma_{kl}  \Phi(\textbf{R})_{kl,i}\Phi(\textbf{R})_{kl,j}R_{kl}$. The remaining non-zero block of $\bf{\Lambda}$ obtained in this way is exact the reduced (induced) metric $\bf{\Lambda}^*$ in the sub-manifold where $\sum_i P_i=1$($\sum_i \dot{P}_i=0$) and $1>P_i>0$ for each i.

\subsection{The dissipation length and the Geodesic equation}

Here, we compare the dissipation length and the thermodynamic length. In fact, in the stochastic thermodynamics, the system which satisfies the instantaneous detailed balance condition $k_{ij}P_i^e=k_{ji}P_j^e$ (which is a specific case of our general setup in the maintext) has the internal energy $u_{i}=-T\text{ln} P_i^e$, then the entropy $S$, the total internal energy $U$ and the free energy $F$ as the function of the system's state $\{P_i(t)\}$ are given by\cite{qian_2010}:
\begin{equation}
    \begin{gathered}S\left[\left\{P_i\right\}\right]=-\sum_i P_i \ln P_i, \quad U\left[\left\{P_i\right\}\right]=\sum_i P_i u_i, \\ F=U-T S=T \sum_i P_i \ln \left(\frac{P_i}{P_i^e}\right) ,\end{gathered}
\end{equation}
where $k_{B}=1$. And the dissipation rate is given by:

\begin{equation}
    e_p(t)=-\frac{1}{T} \frac{d F\left[\left\{P_i(t)\right\}\right]}{d t}=-\frac{1}{T}\sum_i \dot{P}_i \frac{\partial F}{\partial P_i}  =\sum_i \dot{P}_i \ln \left(\frac{P_i^e}{P_i}\right)=\frac{1}{2}\sum_{ij}\left(P_i k_{i j}-P_j k_{j i}\right) \ln \left(\frac{P_i k_{i j}}{P_j k_{j i}}\right),
\end{equation}
compared to the dissipation rate given by Salamon\cite{Geo_1983}: 
\begin{equation}
    e_p(t)=\sum_i\dot{X}_i\left[Y_i^e(t)-Y_i(t)\right] 
 =\sum_{ij} \dot{X}_i \frac{\partial Y_i}{\partial X_j} 
  \left[X_j^e(t)-X_j(t)\right],
\end{equation}
where $X_i$ are extensive variables, $Y_i$ are intensities and the superscript $e$ means the heat bath. Similarly, $\sum_i \dot{P}_i \ln \left(\frac{P_i^e}{P_i}\right)=\sum_i \dot{P}_i (\ln P_i^e - \ln P_i)=\sum_{ij}  \dot{P}_i \eta_{ij}  (P_j^e - P_j)$ where $\eta_{ij} =-\frac{1}{T} \frac{\partial^2 F}{\partial P_i \partial P_j}  =  \frac{1}{P_i} \delta_{i,j}$.  Note that, the thermodynamic length is defined as $\int_0^{\tau}  \sqrt{\sum_{ij}\eta_{ij}\dot{X}_i\dot{X}_j}dt$ where $\eta_{ij} = \frac{\partial Y_i}{\partial X_j} = \frac{\partial^2 U}{\partial X_i \partial X_j}$ and $U(X_i)$ is the internal energy\cite{Geo_1983}. And the statistical distance is defined as $\int_0^\tau \sqrt{\sum_i \frac{1}{P_i}\left[\frac{d P_i}{d t}\right]^2} d t$ by Crooks\cite{Geo_2007}. However, these two lengths do not contain the lag time matrix $\xi_{ij}$, also called the friction matrix in \cite{Geo_2012,OC_2015,OC_2017}, defined by $X_j^e(t)-X_j(t) = \sum_k\xi_{jk} \dot{X}_k$ and $ P_j^e(t)-P_j(t) = \sum_k \xi_{jk} \dot{P}_k$. The entropy production rate is expressed as $e_p(t)=\sum_{ijk}\dot{P}_i\eta_{ik}\xi_{kj}\dot{P}_j=\sum_{ij}\dot{P}_i\Lambda_{ij}\dot{P}_j$ with $\Lambda_{ij}=\sum_k\eta_{ik}\xi_{kj}$, from which we have $\bf{\xi}=\bf{\eta}^{-1}\bf{\Lambda}$ with $\eta^{-1}_{ij}=P_i\delta_{i,j}$.
As a consequence, the dissipation length defined by $\mathcal{L}=\int_0^{\tau}  \sqrt{\sum_{ij}^{N-1}\Lambda^*_{ij}(\textbf{R})\dot{P}_i\dot{P}_j}dt$ from the 2-step optimization scheme, containing the lag time matrix, is different from the thermodynamic length and statistical length. And, we will show in the following section that $\bf{\Lambda^*}$ can not be the Hessian of some effective energy function $\Psi(\bf{P})$ like the thermodynamic length.

With the distance metric, the length of a curve in the probability space is characterized by the dissipation length as $\mathcal{L}=\int_0^{\tau}  \sqrt{\sum_{ij}^{N-1}\Lambda^*_{ij}(\textbf{R})\dot{P}_i\dot{P}_j}dt$(here we use the reduced metric $\bf{\Lambda}^*$ and  $i,j = 1,2,..., N-1$). By using the Cauchy-Shwartz inequality, the totoal dissipation is bounded by the dissipation distance $\mathcal{L}_m$:

\begin{equation}
    \mathcal{C}=\int_0^{\tau} \sum_{ij} \Lambda^*_{ij}(\textbf{R})\dot{P}_i\dot{P}_jdt \geq \big[\int_0^{\tau}  \sqrt{\sum_{ij}\Lambda^*_{ij}(\textbf{R})\dot{P}_i\dot{P}_j}dt\big]^2/{\tau} = \mathcal{L}_m^2/{\tau}, 
\end{equation}
with equality only for $  \sqrt{\sum_{ij}\Lambda^*_{ij}(\textbf{R})\dot{P}_i\dot{P}_j}= const.=\mathcal{L}_m/{\tau}$ which is characterized by the geodesic equation:
\begin{equation}\label{geodesic}
    \ddot{P}_i+\sum_{j k} \mathcal{K}_{j k}^i \dot{P}_j \dot{P}_k=0,
\end{equation}
with the boundary $\bf{P}(0)$ and $\bf{P}(\tau)$. The Christoffel symbol is defined as $ \mathcal{K}_{j k}^i \equiv \frac{1}{2} \sum_l\left(\Lambda^{*-1}\right)_{l i}\left(\partial_{P_k} \Lambda^*_{l j}+\partial_{P_j} \Lambda^*_{l k}-\partial_{P_l} \Lambda^*_{j k}\right).$ The optimal protocol is given by $k_{ij}^*=(P_i+P_j)^{-1} (P_j C_{ij}+\sum_{k}\Phi_{ij,k}\dot{P}_k)$, using the Kirchhoff's Law.

\subsection{Applications to simple cases}
In the following sections, we show how to calculate the metric $\bf{\Lambda}$ step by step with the help of the specific 2-state and 3-state network.

\subsubsection{2-state system}

For the two states system, the conductance matrix is given by:

\begin{equation}
    \bf{\Gamma} = \begin{bmatrix} R_{12}^{-1} & -R_{12}^{-1} \\ -R_{12}^{-1} & R_{12}^{-1} \end{bmatrix},
\end{equation}
where $R_{12}=R_{21}=1/(P_1k_{12}+P_2k_{21})$. Then, wiping out the 2nd row and column of matrix $\bf{\Gamma}$, we get the resulting matrix $\gamma = R_{12}^{-1}$ with dimension 1 and $\bf{\Omega}$ can be obtained by:

\begin{equation}
    \bf{\Omega}=\bf{\Gamma}^{-1} = \begin{bmatrix} R_{12} & 0 \\ 0 & 0 \end{bmatrix},
\end{equation}
followed by the optimal current $J_{12}=\sum_k (\Omega_{1k}-\Omega_{2k})R^{-1}_{12} \dot{P}_k\equiv \sum_k \Phi_{12,k}(\textbf{R})\dot{P}_k=\dot{P}_1$, which is obviously. And the $\bf{\Phi}$ is given by:

\begin{equation}
    \bf{\Phi}= \begin{bmatrix} 1 & 0 \\ -1 & 0 \end{bmatrix},
\end{equation}
and $\bf{\Lambda}$ is 

\begin{equation}
    \bf{\Lambda}= \begin{bmatrix} 2R_{12} & 0 \\ 0 & 0 \end{bmatrix} = \begin{bmatrix}  \bf{\Lambda}^* & 0\\ 0& 0\end{bmatrix},
\end{equation}
followed by $\dot{S}=2\dot{P}_1^2R_{12}=\dot{P}_1^2R_{12}+\dot{P}_2^2R_{12}$, since $\dot{P}_1=-\dot{P}_2$. In the adiabatic limit we have $R_{12}=1/(P_1k_{12}+P_2k_{21}) = \frac{1}{2C_{12}}(\frac{1}{P_1(t)}+\frac{1}{P_2(t)})=\frac{1}{2C_{12}P_1(t)P_2(t)}$, and total dissipation is:

\begin{equation}
    \int_0^{\tau}\dot{S} dt = \int_0^{\tau}\frac{\dot{P}_1^2}{C_{12}P_1(1-P_1)}dt
\end{equation}
of which the Geodesic equation is:
\begin{equation}\label{el}
    \ddot{P}_1 + \frac{2P_1-1}{2P_1(1-P_1)} \dot{P}_1^2=0,
\end{equation}
with $P_1(0)=P_i,P_1(\tau)=P_f$. The solution of Eq.\ref{el} is:
\begin{equation}\label{analy}
    P_1(t) = \text{sin}^2(\frac{1}{2}c_1(t+c_2))
\end{equation}
with $c_1=\frac{2}{\tau}(\text{sin}^{-1}\sqrt{P_f}-\text{sin}^{-1}\sqrt{P_i})$ and $c_2=\frac{\tau \text{sin}^{-1}\sqrt{P_i}}{\text{sin}^{-1}\sqrt{P_f}-\text{sin}^{-1}\sqrt{P_i}}$. The corresponding optimal protocol is given by $k_{21}= \dot{P}_1+C_{12}P_1$. And the minimum dissipation is 
\begin{equation}
\frac{4\left(\sin ^{-1}\sqrt{P_f}-\sin ^{-1}\sqrt{P_i}\right)^2}{ C_{12} \tau}.
\end{equation}

\begin{figure}[htb]
\includegraphics[scale=0.8]{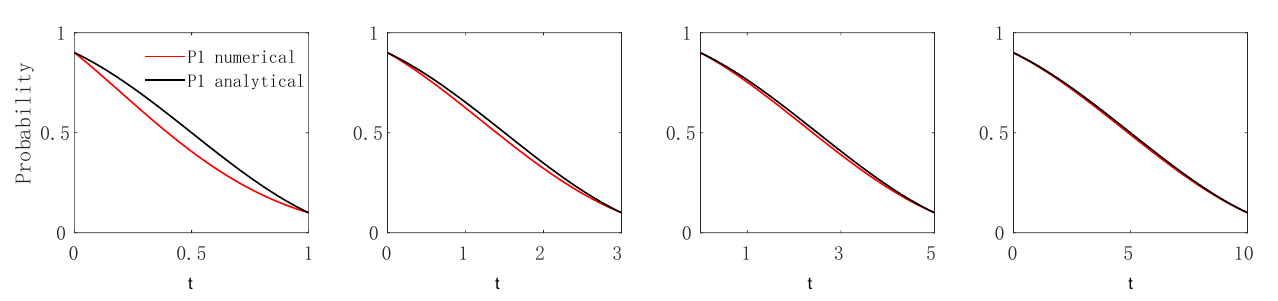}
\caption{\label{P1_tau}The comparison between the numerical solution and analytical solution. As the total control time $\tau$ increases, the system reaches the adiabatic limit and the numerical solution is approaching to the analytical solution obtained in the limit. $P_i = 0.9$, $P_f = 0.1$ and $C_{12}=5$.}
\end{figure}

As shown in Fig.~\ref{P1_tau} , we show the numerical solution (obtained by OC toolbox directly) is approaching the analytical solution Eq.\ref{analy} in the adiabatic limit.


\subsubsection{3-state network}

For the 3-state network, the conductance matrix is given by:

\begin{equation}
    \bf{\Gamma} = \begin{bmatrix} R_{12}^{-1} + R_{13}^{-1} & -R_{12}^{-1} & -R_{13}^{-1} \\ -R_{21}^{-1} & R_{21}^{-1} +R_{23}^{-1} & -R_{23}^{-1} \\ -R_{31}^{-1}& -R_{32}^{-1} & R_{31}^{-1}+R_{32}^{-1} \end{bmatrix},
\end{equation}
where $R_{ij}=1/(P_ik_{ij}+P_jk_{ji})$. Then, wiping out the 3rd row and column of matrix $\bf{\Gamma}$, we get the resulting matrix $\gamma$ with dimension 2 and $\bf{\Omega}$ can be obtained by:

\begin{equation}
\bf{\Omega}=\bf{\Gamma}^{-1}= \begin{bmatrix}
    \frac{R_{13}(R_{12}+R_{23})}{R_{12}+R_{13}+R_{23}} & \frac{R_{13} R_{23}}{R_{12}+R_{13}+R_{23}} & 0 \\
 \frac{R_{13} R_{23}}{R_{12}+R_{13}+R_{23}} & \frac{R_{23} (R_{12}+R_{13})}{R_{12}+R_{13}+R_{23}}& 0\\ 0&0&0 \end{bmatrix},
\end{equation}
followed by:

\begin{equation}
    \bf{\Phi}= \begin{bmatrix} \frac{R_{13}}{R_{12}+R_{13}+R_{23}} & \frac{-R_{23}}{R_{12}+R_{13}+R_{23}} &0\\ \frac{-R_{13}}{R_{12}+R_{13}+R_{23}} &\frac{R_{23}}{R_{12}+R_{13}+R_{23}} & 0\\ \frac{R_{12}+R_{23}}{R_{12}+R_{13}+R_{23}} & \frac{R_{23}}{R_{12}+R_{13}+R_{23}} & 0\\ \frac{-(R_{12}+R_{23})}{R_{12}+R_{13}+R_{23}} & \frac{-R_{23}}{R_{12}+R_{13}+R_{23}}& 0\\ \frac{R_{13}}{R_{12}+R_{13}+R_{23}}& \frac{R_{12}+R_{13}}{R_{12}+R_{13}+R_{23}} &0\\ \frac{-R_{13}}{R_{12}+R_{13}+R_{23}}& \frac{-(R_{12}+R_{13})}{R_{12}+R_{13}+R_{23}} &0

    \end{bmatrix},
\end{equation}
and $\bf{\Lambda}$ is 

\begin{equation}
    \bf{\Lambda}= \begin{bmatrix} 2 \frac{R_{13}(R_{12}+R_{23})}{R_{12}+R_{13}+R_{23}} & 2 \frac{R_{13}R_{23}}{R_{12}+R_{13}+R_{23}} & 0\\ 2 \frac{R_{13}R_{23}}{R_{12}+R_{13}+R_{23}} & 2 \frac{R_{23}(R_{12}+R_{13})}{R_{12}+R_{13}+R_{23}} &0\\ 0& 0&0\end{bmatrix} = \begin{bmatrix}  \bf{\Lambda}^* & 0\\ 0& 0\end{bmatrix},
\end{equation}
followed by $\dot{S}=2(\dot{P}_1^2 \frac{R_{13}(R_{12}+R_{23})}{R_{12}+R_{13}+R_{23}} +\dot{P}_2^2  \frac{R_{23}(R_{12}+R_{13})}{R_{12}+R_{13}+R_{23}} + 2\dot{P}_1\dot{P}_2 \frac{R_{13}R_{23}}{R_{12}+R_{13}+R_{23}})$.  As a consequence of $\dot{P}_3^2 = (\dot{P}_1+\dot{P}_2)^2$, the dissipation rate can be expressed as $\dot{S}=2(\dot{P}_1^2 \frac{R_{13}R_{12}}{R_{12}+R_{13}+R_{23}} +\dot{P}_2^2  \frac{R_{23}R_{12}}{R_{12}+R_{13}+R_{23}} + \dot{P}_3^2 \frac{R_{13}R_{23}}{R_{12}+R_{13}+R_{23}})$ equivalently, with the corresponding new metric:
\begin{equation}
    \tilde{\bf{\Lambda}}= \begin{bmatrix} 2 \frac{R_{13}R_{12}}{R_{12}+R_{13}+R_{23}} & 0 & 0\\ 0 & 2 \frac{R_{23}R_{12}}{R_{12}+R_{13}+R_{23}} &0\\ 0& 0&2\frac{R_{13}R_{23}}{R_{12}+R_{13}+R_{23}}\end{bmatrix}.
\end{equation}

\begin{figure}[htb]
	\centering
	\includegraphics[scale=0.7]{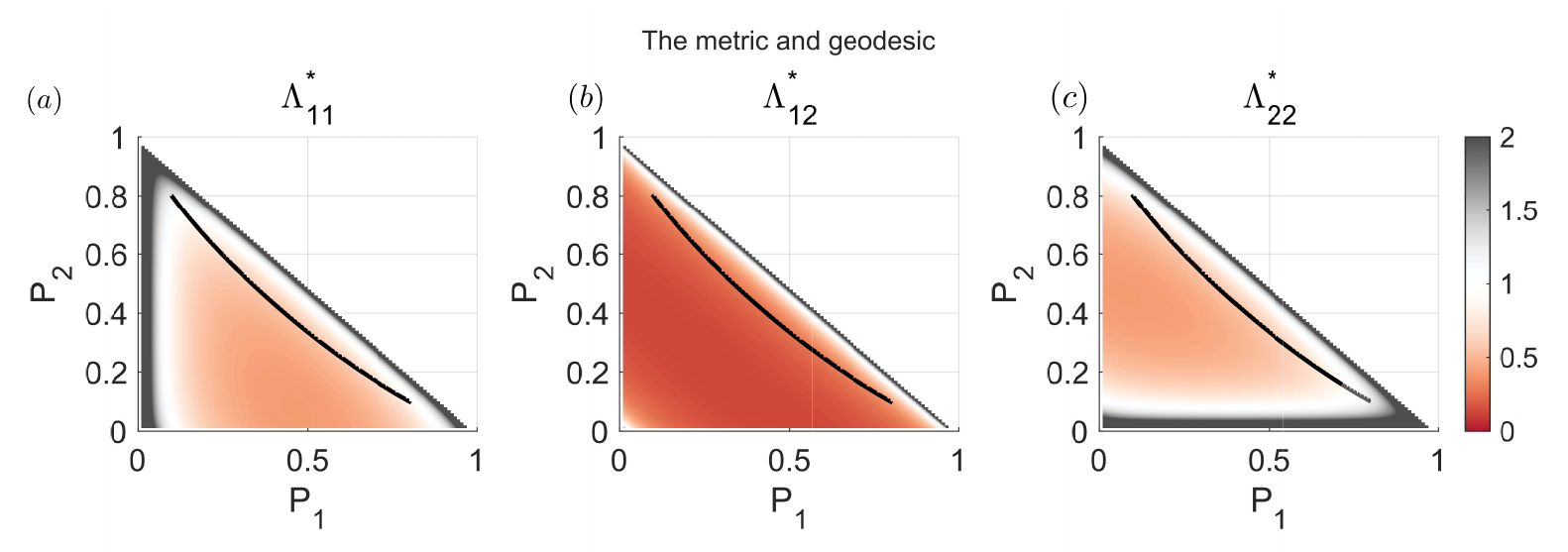}
	\caption{Phase I, the typical parameters set is $C_{12}=5,C_{13}=10,C_{23}=10$. The metric components $\Lambda^*_{ij}$ and geodesic with $P_1(0)=0.8, P_2(0)=0.1, P_1(\tau)=0.1, P_2(\tau)=0.8$ for $C_{12}=5,C_{13}=10,C_{23}=10$. The color represents the value of $\Lambda^*_{ij}$ and the black line is the geodesic. The $\bf{P}$ space is restricted by $P_1>0, P_2>0, P_1+P_2<1$. The off-diagonal elements $\Lambda^*_{12}=\Lambda^*_{21}$ are much smaller than the diagonal elements $\Lambda^*_{11}$ and $\Lambda^*_{22}$, and the projection of arcs of geodesic to axis $P_1$($P_2$) would be larger where $\Lambda^*_{11}$ ($\Lambda^*_{22}$) is smaller. As a consequence, the geodesic bends to the red area in (a) and (c) slightly.}\label{metric1}
\end{figure}

\begin{figure}[htb]
	\centering
	\includegraphics[scale=0.7]{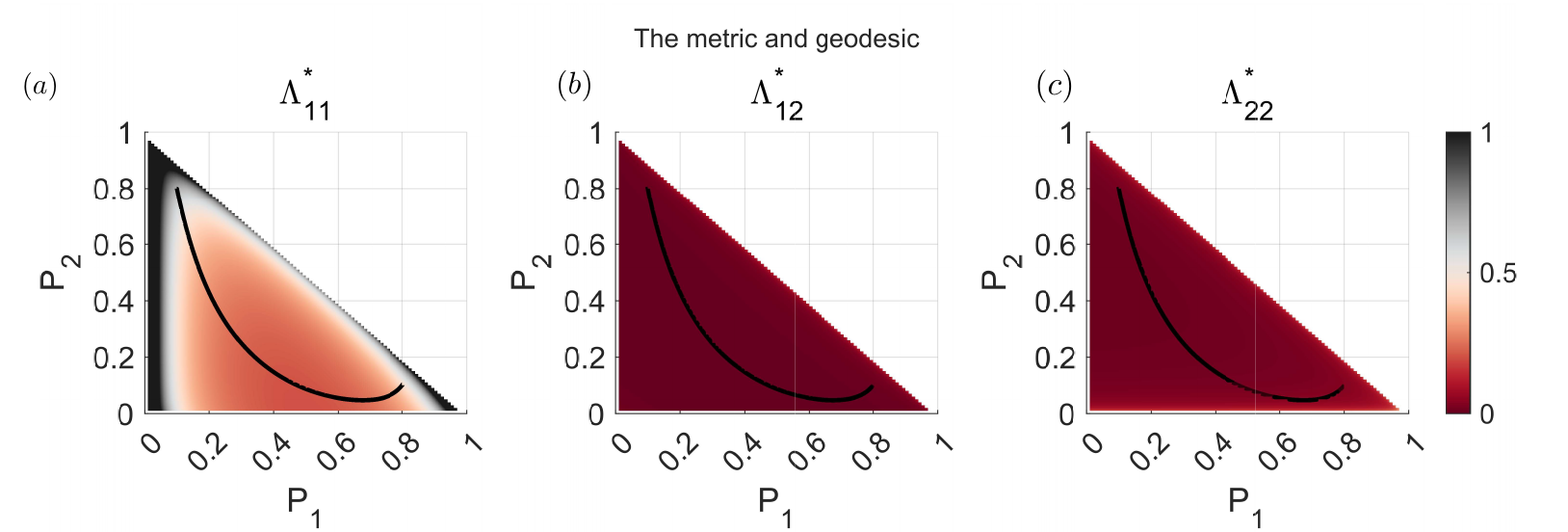}
	\caption{Phase II, the typical parameters set is $C_{12}=5,C_{13}=20,C_{23}=500$. The metric components $\Lambda^*_{ij}$ and geodesic with $P_1(0)=0.8, P_2(0)=0.1, P_1(\tau)=0.1, P_2(\tau)=0.8$ for $C_{12}=5,C_{13}=20,C_{23}=500$. The color represents the value of $\Lambda^*_{ij}$ and the black line is the geodesic. The off-diagonal elements $\Lambda^*_{12}=\Lambda^*_{21}$ and $\Lambda^*_{22}$ are much smaller than $\Lambda^*_{11}$, therefore the the geodesic is dominated by $\Lambda^*_{11}$. As shown in (a), the projection of arcs of geodesic to axis $P_1$ are larger than the projection to axis $P_2$ in the deep red area far away from the diagonal ($P_1+P_2=1$) where $P_3 = 1-P_1-P_2=0$, leading to the indirect geodesic path that goes through a region with high $P_3$. .}\label{metric2}
\end{figure}

\begin{figure}[htb]
	\centering
	\includegraphics[scale=0.7]{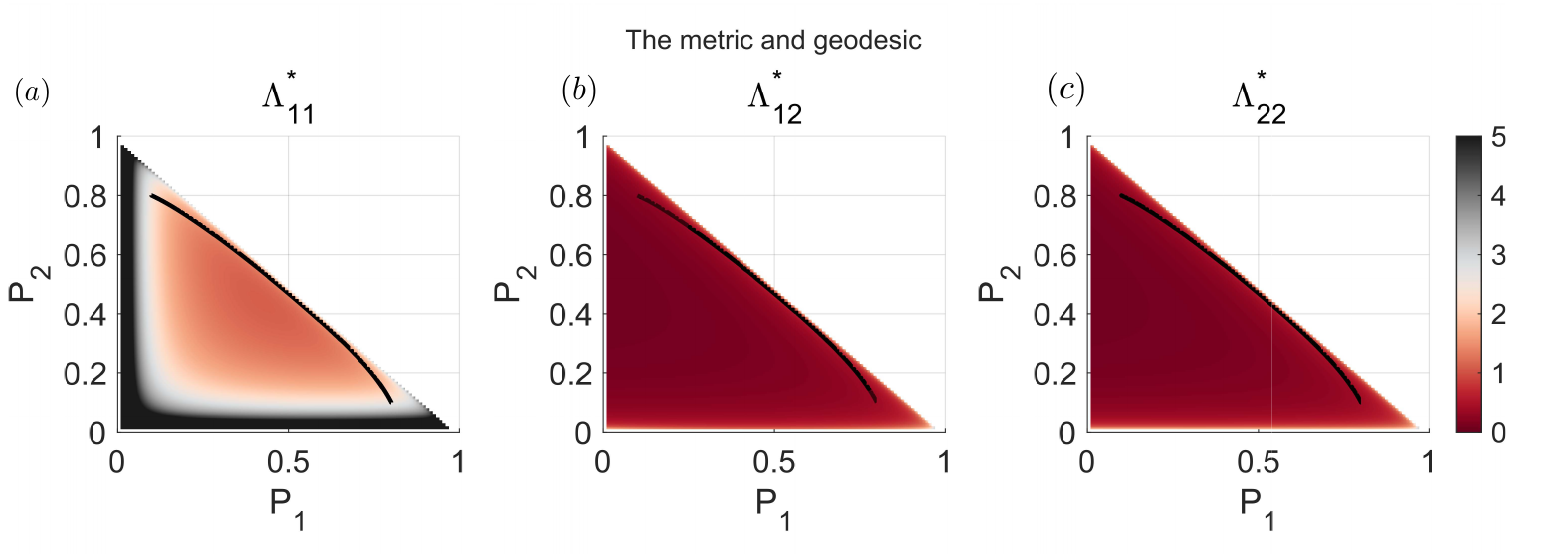}
	\caption{Phase III, the typical parameters set is $C_{12}=5,C_{13}=0.1,C_{23}=50$. The metric components $\Lambda^*_{ij}$ and geodesic with $P_1(0)=0.8, P_2(0)=0.1, P_1(\tau)=0.1, P_2(\tau)=0.8$ for $C_{12}=5,C_{13}=0.1,C_{23}=50$. The color represents the value of $\Lambda^*_{ij}$ and the black line is the geodesic. The off-diagonal elements $\Lambda^*_{12}=\Lambda^*_{21}$ and $\Lambda^*_{22}$ are much smaller than $\Lambda^*_{11}$, therefore the the geodesic is dominated by $\Lambda^*_{11}$. Similarly, the geodesic is inclined to the dark red area near the diagonal ($P_1+P_2=1$) where $\Lambda^*_{11}$ is smaller, leading to a direct path with small $P_3$.}\label{metric3}
\end{figure}

\begin{figure}[htb]
	\centering
	\includegraphics[scale=0.7]{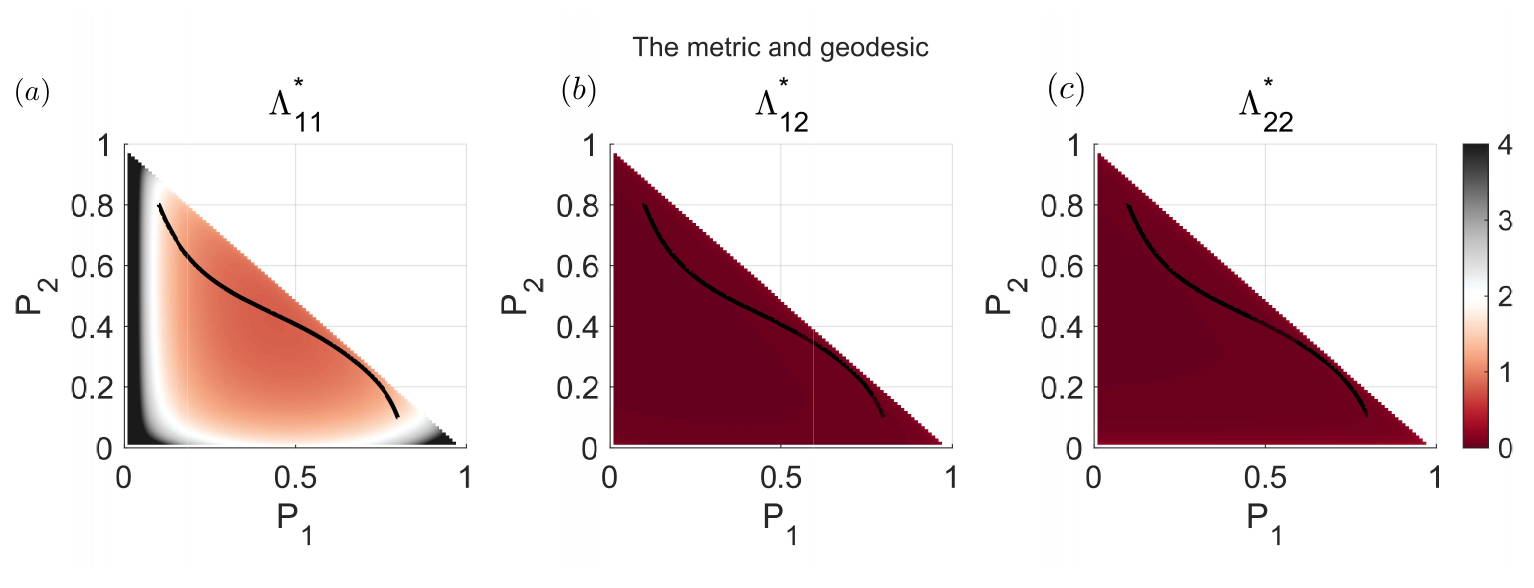}
	\caption{Phase IV, the typical parameters set is $C_{12}=5,C_{13}=2,C_{23}=400$. The metric components $\Lambda^*_{ij}$ and geodesic with $P_1(0)=0.8, P_2(0)=0.1, P_1(\tau)=0.1, P_2(\tau)=0.8$ for $C_{12}=5,C_{13}=2,C_{23}=400$. The color represents the value of $\Lambda^*_{ij}$ and the black line is the geodesic. The off-diagonal elements $\Lambda^*_{12}=\Lambda^*_{21}$ and $\Lambda^*_{22}$ are much smaller than $\Lambda^*_{11}$, therefore the the geodesic is dominated by $\Lambda^*_{11}$. Different form Phase II and III, the basin of red region is slightly away from the diagonal and the geodesic shows a ``S"-shape bending. }\label{metric4}
\end{figure}

Depending on the strength of links $C_{ij}$, the geodesics with metric tensor are shown in Fig.~\ref{metric1}-\ref{metric4} for the 4 typical Phases characterized in Fig.~\ref{md1}.

Herein, we show that $\bf{\Lambda}^*$ is not the hessian of some effective energy function $\Psi(\bf{P})$ by using the case of 3-state network. 
The third mixed partial derivatives is given by $\frac{\partial \Lambda^*_{11}}{P_2}$ and $\frac{\partial \Lambda^*_{12}}{P_1}$:

\begin{equation}
    \begin{aligned}
       \frac{\partial \Lambda^*_{11}}{P_2}&= \frac{-C_{12}^2 (P_1-1) \left(C_{23} (P_2-1)^2 (P_1+2 P_2-1)-C_{13} (P_1-1) P_1^2\right)}{(P_1+P_2-1)^2 \left(C_{12} (C_{13} (P_1-1) P_1+C_{23} (P_2-1) P_2)+C_{13} C_{23} \left(P_1^2+P_1 (2 P_2-1)+(P_2-1) P_2\right)\right)^2}\\&
        -\frac{C_{12} C_{23} (P_1+P_2-1) \left(C_{23} (P_2-1)^2 (P_1+P_2-1)-2 C_{13} (P_1-1) P_1 (P_1+P_2)\right)}{(P_1+P_2-1)^2 \left(C_{12} (C_{13} (P_1-1) P_1+C_{23} (P_2-1) P_2)+C_{13} C_{23} \left(P_1^2+P_1 (2 P_2-1)+(P_2-1) P_2\right)\right)^2}\\&
        +\frac{C_{13} C_{23}^2 \left(P_1^2+P_1 (2 P_2-1)+(P_2-1) P_2\right)^2}{(P_1+P_2-1)^2 \left(C_{12} (C_{13} (P_1-1) P_1+C_{23} (P_2-1) P_2)+C_{13} C_{23} \left(P_1^2+P_1 (2 P_2-1)+(P_2-1) P_2\right)\right)^2},
    \end{aligned}
\end{equation}

\begin{equation}
    \begin{aligned}
        \frac{\partial \Lambda^*_{12}}{P_1}&=\frac{C_{12} (P_2-1) C_{12} \left(C_{23} (P_2-1) P_2^2-C_{13} (P_1-1)^2 (2 P_1+P_2-1)\right)}{(P_1+P_2-1)^2 \left(C_{12} (C_{13} (P_1-1) P_1+C_{23} (P_2-1) P_2)+C_{13} C_{23} \left(P_1^2+P_1 (2 P_2-1)+(P_2-1) P_2\right)\right)^2}
        \\&+\frac{C_{13} C_{23} \left(-2 P_1^3+P_1^2 (5-3 P_2)+P_1 (6 P_2-4)+P_2^3+P_2^2-3 P_2+1\right)}
        {(P_1+P_2-1)^2 \left(C_{12} (C_{13} (P_1-1) P_1+C_{23} (P_2-1) P_2)+C_{13} C_{23} \left(P_1^2+P_1 (2 P_2-1)+(P_2-1) P_2\right)\right)^2}.
    \end{aligned}
\end{equation}
with which, we have: 
\begin{equation}
\begin{aligned}
    \frac{\partial \Lambda^*_{12}}{P_1} - \frac{\partial \Lambda^*_{11}}{P_2}&= \frac{C_{12}^2 \left(C_{13} (P_1-1)^2-C_{23} (P_2-1)^2\right)+C_{13} C_{23}^2 (P_1+P_2)^2}{\left(C_{12} (C_{13} (P_1-1) P_1+C_{23} (P_2-1) P_2)+C_{13} C_{23} \left(P_1^2+P_1 (2 P_2-1)+(P_2-1) P_2\right)\right)^2}\\&-\frac{C_{12} C_{23} \left(C_{13} \left(-2 P_1^2-2 P_1 P_2+2 P_1+P_2^2+2 P_2-1\right)+C_{23} (P_2-1)^2\right)}{\left(C_{12} (C_{13} (P_1-1) P_1+C_{23} (P_2-1) P_2)+C_{13} C_{23} \left(P_1^2+P_1 (2 P_2-1)+(P_2-1) P_2\right)\right)^2} \\&\neq 0,
\end{aligned}
\end{equation}
which means the third mixed partial derivatives
are not the same, as a consequence, $\Lambda^*$ can not be the hessian of some function $\Psi(\bf{P})$.

\section{Comparing the Kirchhoff's bound with the Wasserstein based bound}

Based on the wasserstein distance on graph $\mathcal{W}_1\left(p(0), p(\tau)\right)$, the state-of-the-art thermodynamic bound $\Sigma_\tau = \mathcal{W}_1\left(p(0), p(\tau)\right)^2/\bar{D} \tau$, or known aS the thermodynamic speed limit\cite{SL_2018,OT_2022,OT_2023}, gives the minimum total energy dissipation for fixed time averaged dynamical state mobility $\bar{D}:=\frac{1}{\tau} \int_0^\tau m(t)dt, m = \sum_{i>j} J_{ij}/(\ln P_ik_{ij}-\ln P_ik_{ji})$. By using the inequality $(a-b)/(lna-lnb) \leq (a+b)/2$, the dynamical state mobility is upper bounded by the global reaction activities $\bar{D} \leq \bar{A}$, where the equality holds in the slow driving limit and $\bar{A}:=\frac{1}{\tau} \int_0^\tau d t A(t), A = \sum_{i>j}P_ik_{ij} = \sum_{i>j}(2R_{ij})^{-1}$\cite{OT_2023}. 

Naturally, with different constrains on the transition rate space, $\Sigma_\tau$ gives a lower bound for our setup as fixed $C_{ij}$. Specifically, for a set of $C_{ij}$, there exists an optimal protocol and $\textbf{P}^*(t)$, then one can calculate the corresponding $\bar{D}$ or $\bar{A}$ with the lower bound $\Sigma_\tau$.

As compare to the Kirhhoff's bound, we get rid of the $P(t)$ dependence of $\bar{A}$ by using the same inequality  $R_{ij}>2/C_{ij}$, and get the wasserstein based bound $\Sigma_{wst}$:

\begin{equation}\label{W}
    \mathcal{C}  \geq \mathcal{W}_1\left(p(0), p(\tau)\right)^2/\bar{A} \tau \\> 2\frac{\lVert \textbf{P}(0)-\textbf{P}(\tau)\rVert^2}{  ( C_{12}  + C_{13} + C_{23})\tau},
\end{equation}
where $\mathcal{W}_1\left(p(0), p(\tau)\right)=  \Sigma_i |\Delta P_i|/2$ for fully connected network. Especially, it can be expressed as $\lVert \textbf{P}(0)-\textbf{P}(\tau)\rVert /\sqrt{2}$ in the case of $\Delta P_3 = 0$.

\begin{figure}[!ht]
    \centering
    \includegraphics[scale = 0.5]{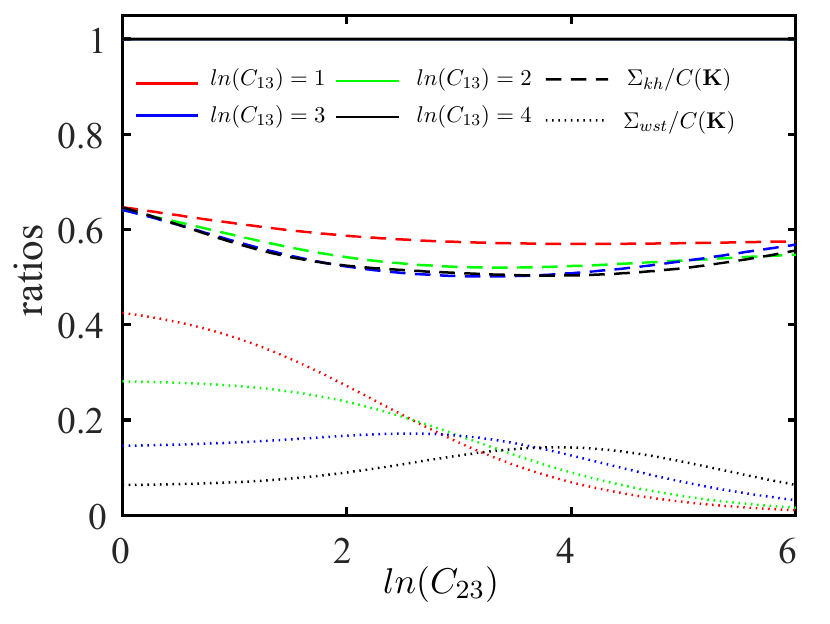}
    \caption{The ratios between the two lower bounds $\Sigma_{hk}$ (dashed lines), $\Sigma_{wst}$ (dotted lines) and the exact minimum total entropy production. The numerical solution for different set of network parameters are selected from several horizontal lines in Fig.~\ref{md1}. Different colors represent different value of fixed $C_{13}$ for fixed $\tau=5, C_{12}=5$.}    \label{boundplt}
\end{figure}

The Kirchhoff's bound in this 3-state network case is 
\begin{equation}
    \Sigma_{kh}=2 
    \frac{\lVert \textbf{P}(0)-\textbf{P}(\tau)\rVert^2}{(C_{12} + C_{13} C_{23}/(C_{13}+C_{23}) )\tau}.
\end{equation} Comparing these two lower bound $\Sigma_{kh}$ and $\Sigma_{wst}$, in the sense of circuit theory, $\Sigma_{wst}$ results from connecting all the channels 1-2, 1-3 and 3-2 in parallel regardless of the structure of the network, different from the bound $\Sigma_{kh}$ derived from the Kirchhoff's law that does contain the topological information of the CRN. As a consequence, $\Sigma_{kh}$ from the Kirchhoff's law is tighter (better) than $\Sigma_{wst}$ as shown in Fig.~\ref{boundplt}.

Note that, beyond this specific 3-state problem, with the topological information, the Kirchhoff's bound $\Sigma_{kh}$ would be generally more meaningful and better than the bound $\Sigma_{wst}$ for larger CRNs or different $\textbf{P}(0)$ and $\textbf{P}(\tau)$ .

\section{Optimal protocols in adiabatic limit and with finite control time}
\subsection{The OC Phase-diagram in adiabatic limit}

To illustrate the key role of $P(t)$-dependence of $R_{ij}(t)$ in the optimal transport process, we show the numerical solutions of the 3-state transport problem shown in the main text for different fixed channel capacities $\{C_{ij}\}$ with $C_{ij}=k_{ij}+k_{ji}$ and $R_{ij}(t) \approx \frac{1}{2C_{ij}}(\frac{1}{P_i(t)}+\frac{1}{P_j(t)})$ in the adiabatic limit. And the controllable input are the ratios of forward and backward reaction rates $r_{ij}=\frac{k_{ij}}{k_{ji}}$. Since the discontinuous jump steps at the start and the end of the control process are allowed, the reaction rates before
and after the control process $t<0,t>\tau$ can be replaced by the rates $k_{ij}$ that satisfy the detailed balance condition $P_i(0)k_{ij}(t<0)=P_j(0)k_{ji}(t<0), P_i(\tau)k_{ij}(t>\tau)=P_j(\tau)k_{ji}(t>\tau)$, subject to the local constrain 
$C_{ij}=k_{ij}+k_{ji}$.

In the adiabatic limit, the solutions can be obtained by solving the Geodesic equation or using the numerical toolbox for Optimal control problem. Note that the solution obtained by using the OC toolbox directly maybe slightly different from solving the geodesic equation Eq.\ref{geodesic} that can not characterize the jump steps due to the continuous assumption in the adiabatic limit. The solutions showed below are from the numerical toolbox for OC problem directly.

\begin{figure}[htb]
\includegraphics[scale=0.8]{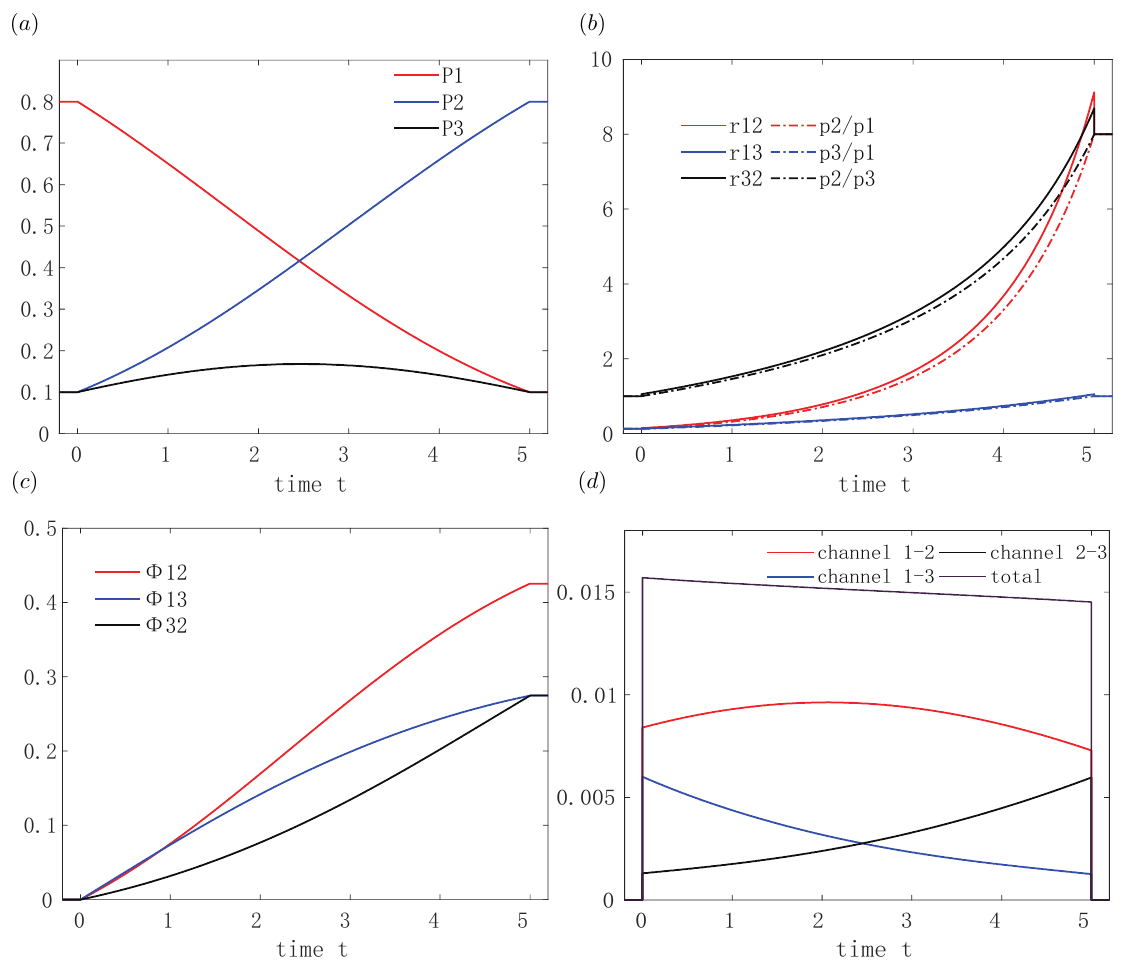}
\caption{\label{Phase1}Phase I. Optimal protocol and trajectory given $C_{12}=5,C_{13}=10,C_{23}=10$, where $C_{12}=C_{13}C_{23}/(C_{13}+C_{23})$. The dimensionless control time window is $0\leq \tau \leq 5$, where the adiabatic limit condition holds.  a) The optimal state trajectory, where $P_1$ and $P_2$ change almost linearly and $P_3$ exhibits an accumulation of probability that reduces the $(\frac{1}{P_1}+\frac{1}{P_3})$-dependent $R_{13}$ and minimizes the cost. b) The solid lines represent the optimal protocol of control input $r_{12},r_{13},r_{32}$, and the dashed lines represent the state of system that always go after the control input with a relaxation distance. Jump steps of control input occur at the start and end of the control process. c) The accumulated probability flow on each reaction channel with no change of flow direction. The total probability flow through channel 1-2 and channels 1-3-2 are nearly equally divided since their ``effective capacities" are equal. d) The entropy production rates (epr) in each channels. The total entropy production rate is nearly a constant.}
\end{figure}

\begin{figure}[htb]
\includegraphics[scale=0.8]{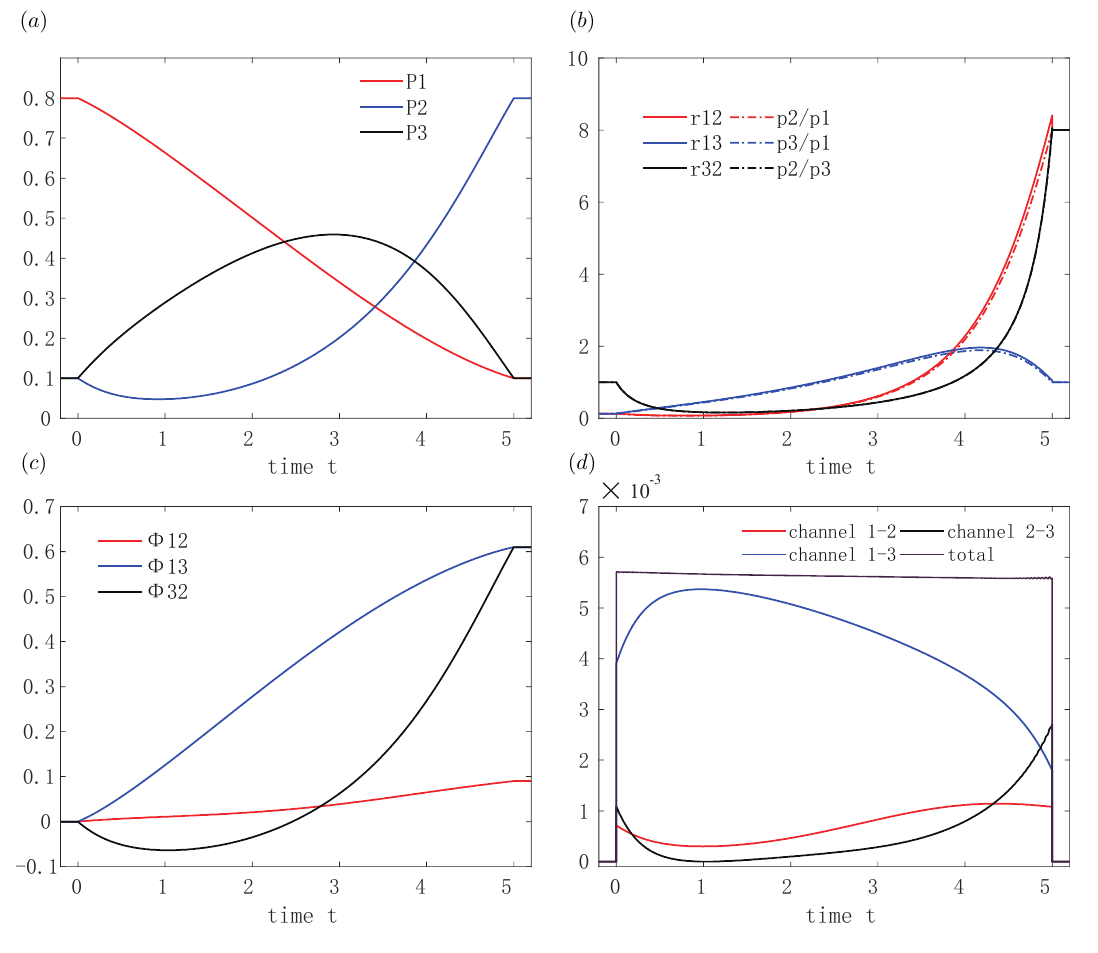}
\caption{\label{Phase2}Phase II. Optimal protocol and trajectory given $C_{12}=5,C_{13}=20,C_{23}=500$, where $C_{12}\ll C_{13}C_{23}/(C_{13}+C_{23})$. a) The optimal state trajectory. different from Phase I, $P_2$ decreases first, then increases, leading to a lager accumulation of $P_3$. b) The optimal protocol. Due to the large $C_{23}$ on channel 3-2, the solid line and the dashed line almost coincide, indicating that the reaction on this channel balances quickly. c) The probability flow mainly passes through channels 1-3-2, because the effective capacity in channels 1-3-2 is larger than 1-2. Part of probability initially flows from state 2 towards state 3, causing a decrease in $P_2$ at the initial moment and more probability accumulation on $P_3$. This is attributed to the fact that $C_{23}$ is much larger than $C_{13}$, resulting in a lower-cost transport on channel 2-3, which contributes to the reduction of entropy production on channel 1-3 by absorbing the probability from state 2 to state 3. d) Due to the large $C_{23}$, entropy production rates in channel 1-2 and 2-3 are small.} 
\end{figure}

\begin{figure}[htb]
\includegraphics[scale=0.8]{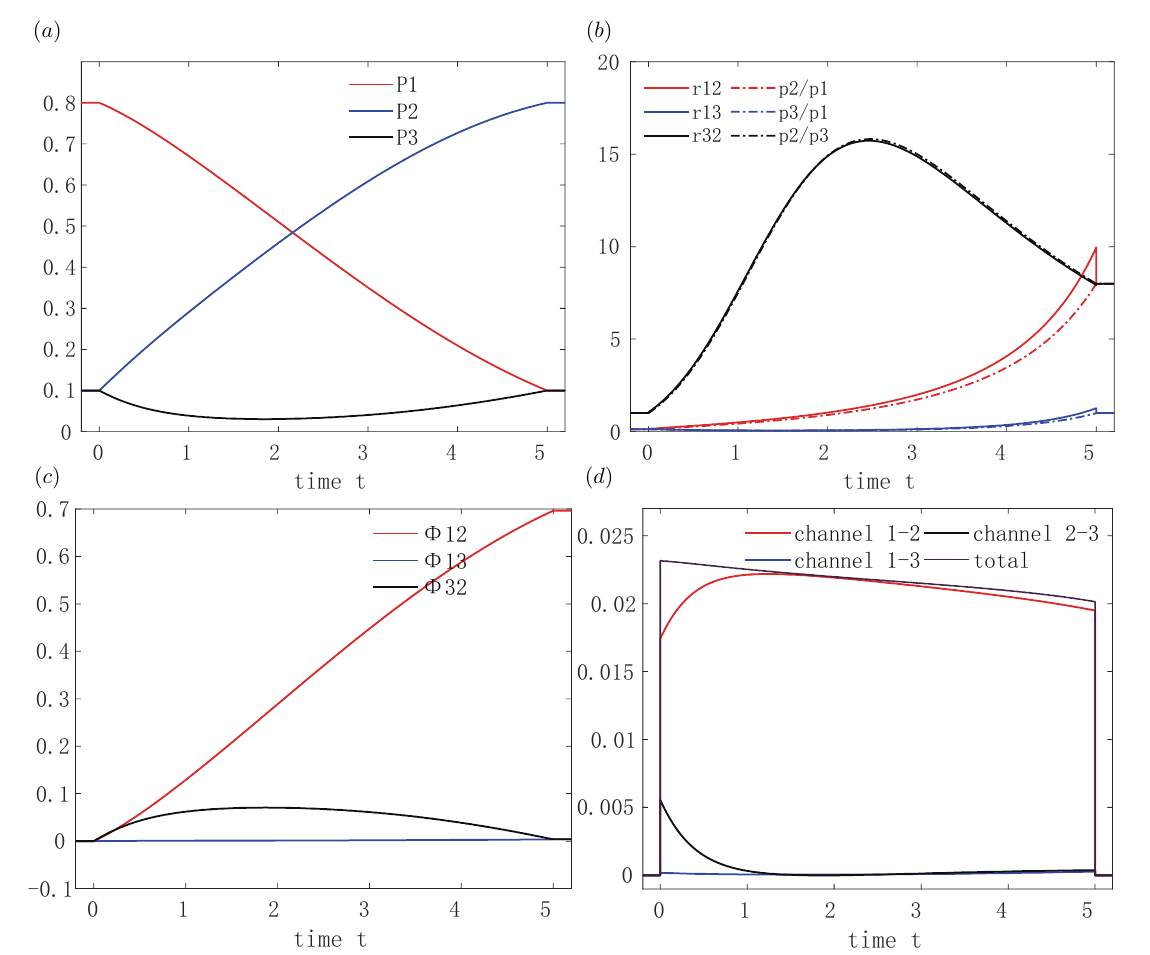}
\caption{\label{Phase3}Phase III. Optimal protocol and trajectory given $C_{12}=5,C_{13}=0.1,C_{23}=50$, where $C_{12}\gg C_{13}C_{23}/(C_{13}+C_{23})$. a) The optimal state trajectory, where P1 and P2 nearly undergo linear changes, while P3 experiences a process of fast decreasing and then recover. b) The optimal protocol, where $r_{32}$ and $p_2/p_3$ first increase, then decrease, corresponding to the trajectory of $P_3$. Similarly, due to the high value of $C_{23}$ on channel 3, the solid and dashed lines almost coincide.  c) Almost all probability flux passes through channel 1-2, because the effective capacity in channels 1-2 is much larger than 1-3-2. Initially, most of probability in state 3 flows towards state 2 first to reduce the cost in channel 1-2 then flows back to state 3. d) Entropy production rates.}
\end{figure}

\begin{figure}[htb]
\includegraphics[scale=0.8]{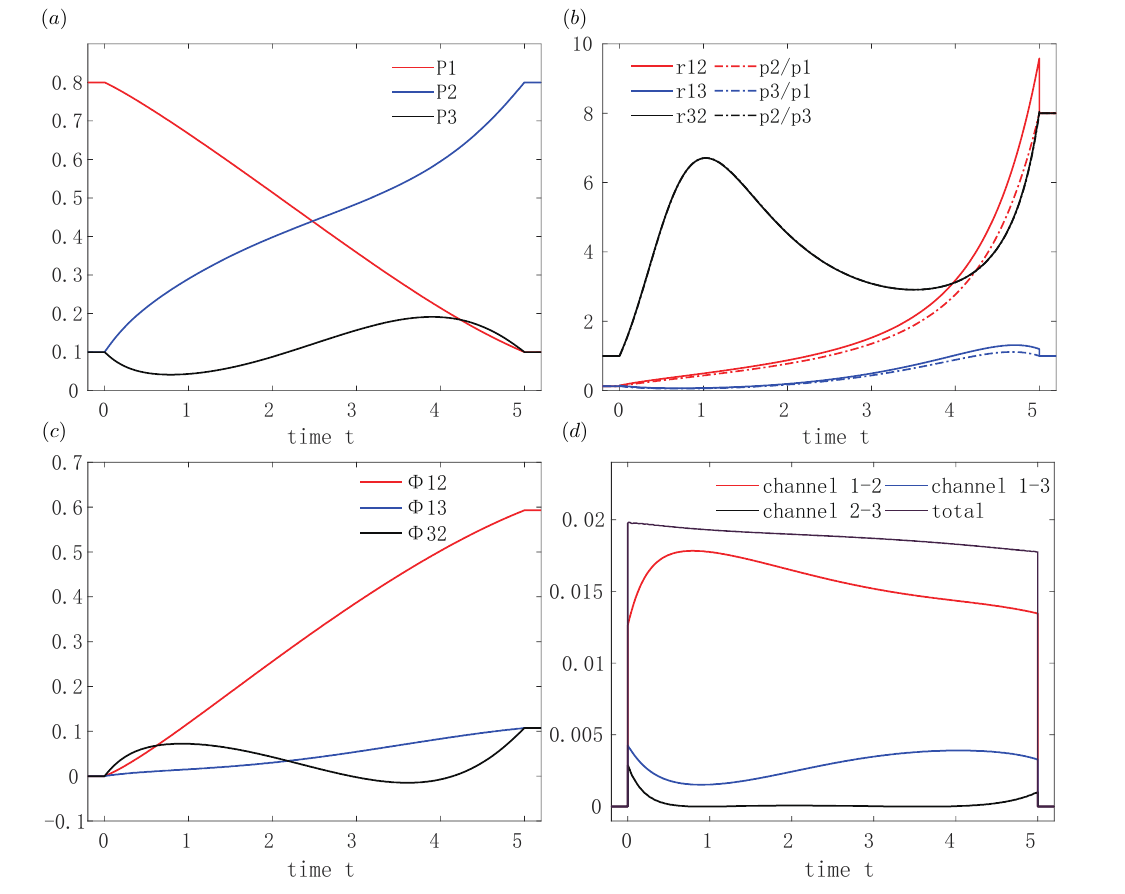}
\caption{\label{Phase4}Phase IV. Optimal protocol and trajectory given $C_{12}=5,C_{13}=2,C_{23}=400$, where $C_{12}>C_{13}C_{23}/(C_{13}+C_{23})$. a) The optimal state trajectory, where P1 and P2 nearly undergo linear changes, while P3 experiences a process of first decreasing, then accumulating, and finally releasing. b) The optimal protocol, where $r_{32}$ and $p_2/p_3$ first increase, then decrease and finally increase, corresponding to the trajectory of $P_3$. Similarly, due to the high value of $C_{23}$ on channel 3-2, the solid and dashed lines almost coincide. c) The probability flow mainly passes through channel 1-2, because the effective capacity in channels 1-3-2 is larger than 1-2. Initially, most of probability in state 3 flows towards state 2 first to reduce the cost in channel 1-2 since $C_{12}$ is larger. When $P_2$ becomes large enough, the flow in channel 3-2 goes back to accumulate in state 3 to reduce the cost on channel 1-3. Finally, it redirects again towards state 2 to reach the targeted distribution. d) Due to the large $C_{23}$, entropy production rates in channel 12 and 23 are small.}
\end{figure}

In the main text, we show the coarse-grained Phase diagram based on the percentage of transport
going through the direct link 1-2 versus the indirect route 1-3-2. Herein, taking the $P(t)$-dependence of $R_{ij}(t)$ into consideration, numerical solution shows 4 kinds of distinct transport phases characterized by the shape of optimal $\textbf{P}^*(t)$ trajectories, each of which corresponds to a specific set of capacities $C_{12},C_{13},C_{23}$ as shown in Fig.~\ref{Phase1}-\ref{Phase4}, where we show the optimal $\textbf{P}^*(t)$ trajectory and the optimal control input. In addition, we record the accumulated flux $\Phi_{ij}(t) = \int_{0}^{t} J_{ij} dt$ to monitor the currents in each channel. And we also show the entropy production rate for each channels.

\begin{figure}[htb]
	\centering
	\includegraphics[scale=1.2]{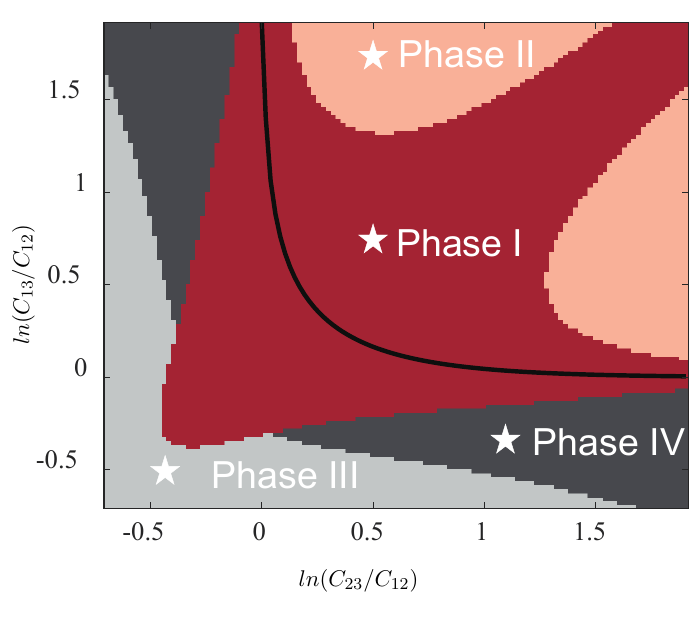}
	\caption{OC Phase diagram for fixed $C_{12}=5$ and dimensionless control time $\tau = 5$ with $C_{23},C_{13}$ as parameters. Due to the system's symmetry, the phase diagram is mirror-symmetric along the axis where $C_{13}=C_{23}$. The system exhibits four different phases: Phase I corresponds to the situation depicted in Fig.~\ref{Phase1}; Phase II corresponds to Fig.~\ref{Phase2}; Phase III corresponds to Fig.~\ref{Phase3}; Phase IV corresponds to Fig.~\ref{Phase4}; Different phases are separated by the number of times of flux redirection in channel 32: Phase I has zero time of redirection compared to Phase II, which has one; Phase III has two times of redirection compared to Phase I; Phase IV has one time of redirection compared to Phase III. The black line is the line where $C_{12}= C_{13}C_{23}/(C_{13}+C_{23})$.}\label{md1}
\end{figure}

The intrinsic feature of these various phases of transport is the direction of probability flow on reaction channel 3-2. Generally, the direction remains consistent throughout, but under certain conditions, the flow direction reverses in order to reduce the dissipation. These 4 phases of transport are delineated within the parameter space, leading to the Phase-Diagram Fig.~\ref{md1}.

For Phase I shown in Fig.~\ref{Phase1}, the typical parameters set is $C_{12}=5,C_{13}=10,C_{23}=10$, from which we have $C_{12}=C_{13}C_{23}/(C_{13}+C_{23})$. Therefore, the total probability flow through the direct channel 1-2 and indirect channels 1-3-2 are nearly equally divided. And $P_3$ exhibits an accumulation of probability that reduces the $(\frac{1}{P_1}+\frac{1}{P_3})$-dependent $R_{13}$ to minimize the cost.

For Phase II shown in Fig.~\ref{Phase2}, the typical parameters set is $C_{12}=5,C_{13}=20,C_{23}=500$, from which we have $C_{12}<C_{13}C_{23}/(C_{13}+C_{23})$. Therefore, most of probability goes through the indirect channels 1-3-2, and the cost of transport in channel 3-2 is cheaper because of large $C_{23}$, part of probability in state 2 initially flows towards state 3, decreasing $P_2$ at the initial moment, leading to more probability accumulation on $P_3$ to help reduce the cost in channel 1-3. 

For Phase III shown in Fig.~\ref{Phase3}, the typical parameters set is $C_{12}=5,C_{13}=0.1,C_{23}=50$, from which we have $C_{12}\gg C_{13}C_{23}/(C_{13}+C_{23})$. Therefore, most of probability goes through channel 1-2 directly. As a result of large $C_{23}$, the cost of transport in channel 3-2 is cheaper. As a consequence, most of probability in state 3 flows towards state 2 to help reduce $R_{12}$ initially, leading to a lower cost in channel 1-2. As time goes, that part of assisted probability would flows back to state 3 to reach the targeted distribution.

For Phase IV shown in Fig.~\ref{Phase4}, the typical parameters set is $C_{12}=5,C_{13}=2,C_{23}=400$, from which we have $C_{12}> C_{13}C_{23}/(C_{13}+C_{23})$. Note that this Phase is the immediate Phase between Phase I and III while $C_{13}$ decreases and $C_{23}$ increases as shown in phases diagram Fig.~\ref{md1}. Initially, with large $C_{23}$, most of probability in state 3 flows towards state 2 to help reduce $R_{12}$. As time goes, the probability in state 2 is big enough but the probability in state 3 is too small. Along with a considerable flux goes through channel 1-3-2 different from Phase III, the probability will flow back to state 3 through channel 2-3 to reduce the cost in channel 1-3.

The existence of jump steps at the start and end of optimal protocols has been identified\cite{OC_2007,OC_2021} with the discontinuously changed entropy production rate. In the slow driving (adiabatic) limit, the jump step is approaching zero and the optimal entropy production rate along the geodesic in $\textbf{P}$ space is constant.

\subsection{Optimal control with finite control time $\tau$}

Here, we show that the shortest control
time $\tau_c$ within which the desired change of probability can
be achieve is of the same order to the relaxation time $\tau_R$: $\tau_c/\tau_R=\mathcal{O}(1)$. 

For the general case in CRN, the shortest control
time $\tau_c$ is dependent on the topology of network and the strength $C_{ij}$ for each links. The CME can be written in the matrix form:
\begin{equation}
    \dot{\bf{P}}=\bf{K}\bf{P},
\end{equation}
where $K_{ij} =k_{ji}$ for $i\neq j$ and $K_{ii}= -\Sigma_j k_{ij}$. And the smallest minus non-zero eigenvalue $\lambda^*(t)$ of $\bf{K}$  determines the averaged relaxation time $\tau_R = \frac{1}{\tau_c}\int_0^{\tau_c} \frac{1}{\lambda^*(t)} dt$.

Considering the control process, the fastest transport path from node $m$ to $n$, $m\rightarrow \cdots \rightarrow k \rightarrow \cdots \rightarrow n$, would prefer the links those have relative large $C_{ij}$ which determines the shortest control time $\tau_c$. Simultaneously, the OC protocol also prefers the path with links of large $C_{ij}$, which determines the relaxation time $\tau_R$, intuitively leading to the relation: $\tau_c/\tau_R=\mathcal{O}(1)$.

As an illustration, we take the 2-state transport for example, where the master equation is:
\begin{equation}\label{2state}
    \frac{d P_1}{d t}= P_2 k_{21} -P_1 k_{12}=k_{21} -P_1C_{12},
\end{equation}
with the boundary condition: $P_1(0)=P_i>P_1(\tau)=P_f$. The shortest control time $\tau_c$ is obtained when $k_{21}=0$ and the solution of the above equation is $P_1(t) = P_i \text{e}^{-C_{12}t}$, followed by $\tau_c=\ln\frac{P_i}{P_f}/C_{12}$. 

For the linear system governed by the CME, the relaxation time can be obtained near equilibrium by using the linear response approximation where the probability $P_{1}(t)$ can be written as the quasiequilibrium value $P_{a}=\frac{k_{21}}{k_{12}+k_{21}}$ plus a small correction $\delta P$ :
\begin{equation}\label{eq1}
P_{1}(t)=\frac{k_{21}}{k_{12}+k_{21}}+\delta P(t),
\end{equation}
which can be plugged into two states CME to obtain:
\begin{equation}\label{eq2}
\frac{d P_{a}}{d t}+\frac{d \delta P}{d t}=-\left(k_{12}+k_{21}\right) \delta P.
\end{equation}
Supposing that $k_{ij}$ is fixed, we have $\delta \dot{P}(t) = -C_{12} \delta P(t)$, therefore, the relaxation time scale is characterized by $1/C_{12}$ and we have $\tau_c/\tau_R=\mathcal{O}(1)$.

For the 3-state network model considered in the main text, in the case of direct transport where $C_{12}\gg C_{13}C_{23}/(C_{13}+C_{23})$, the preferred path is 1-2 and we have $\tau_c/\tau_R=\mathcal{O}(1)$ the same as the 2-state case. 

In the case of indirect transport 1-3-2 where $C_{12}\ll C_{13}C_{23}/(C_{13}+C_{23})$, the fastest transport is governed by:

\begin{equation}\label{fast-3}
    \begin{cases}
        \dot{P_1}=-C_{13}P_1\\
        \dot{P_3}=C_{13}P_1-C_{23}P_3\\
        \dot{P_2}=C_{23}P_3
    \end{cases}
\end{equation}
with the boundary condition (P1 = 0.8, P2 = P3 = 0.1)
at t = 0 and (P2 = 0.8, P1 =
P3 = 0.1) at t = $\tau_c$. From the Eq.\ref{fast-3}, $\tau_c$ is of order $\mathcal{O}(1)/C_{13}$ if $C_{13}\ll C_{23}$, or $\mathcal{O}(1)/C_{23}$ if $C_{13}\gg C_{23}$, or $\mathcal{O}(1)/C_{13} + \mathcal{O}(1)/C_{23}$ if $C_{13} \sim C_{23}$. 
The relaxation is characterized by the eigenvalue of the transition matrix which in this case is given by:

\begin{equation}
    \bf{K}=\begin{bmatrix}
        -k_{13}& 0& k_{31}\\ 0& -k_{23}& k_{32}\\k_{13}& k_{23}&-k_{32} - k_{31}
    \end{bmatrix},
\end{equation}
of which the smallest minus non-zero eigenvalue is $\frac{-(C_{13}+C_{23})+\sqrt{(C_{13}-C_{23})^2+4k_{31}k_{32}}}{2}$ with $k_{ij}\approx(P_i+P_j)^{-1} P_j C_{ij}$ in the linear response region. Therefore, by keeping the first order term in $\mathcal{O}(\frac{C_{13}}{C_{23}})$ ($\mathcal{O}(\frac{C_{23}}{C_{13}})$) of the expansion of the eigenvalue, it can be calculated that $\tau_R$ is of order $\mathcal{O}(1)/C_{13}$ ($\mathcal{O}(1)/C_{23}$) if $C_{13}\ll C_{23}$ ($C_{13}\gg C_{23}$). If $C_{13} \sim C_{23}$, $\tau_R$ is supposed to be of order $\mathcal{O}(1)/C_{13} + \mathcal{O}(1)/C_{23}$. As a consequence, we have $\tau_c/\tau_R=\mathcal{O}(1)$. 

In the case where $C_{12} \sim C_{13}C_{23}/(C_{13}+C_{23})$, the indirect path 1-3-2 can be considered as an effective direct path 1$-^*$2 with effective $\tilde{C}_{12}
\approx C_{13}C_{23}/(C_{13}+C_{23})$, together with the true direct path 1-2, these two paths can be considered as one direct path with $C^*_{12}=\tilde{C}_{12}+C_{12}$. 
Resulting from the 2-state case, $\tau_c$ can be approximately given by $\ln \frac{P_1(0)}{P_1(\tau)}/(C_{12}+C_{13}C_{23}/(C_{13}+C_{23}))$.  

In the following, we show that the OC protocol evolves with the control time $\tau$. The optimal protocols beyond the slow driving limit can be obtained by the numerical toolbox only.

\begin{figure}[htb]
\includegraphics[scale=1.5]{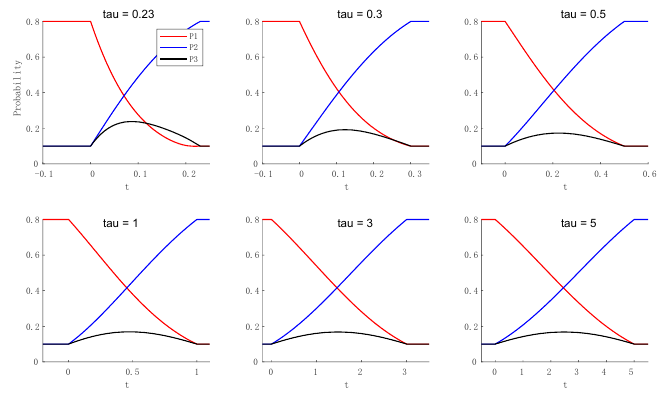}
\caption{Phase I: $C_{12}=5,C_{13}=10,C_{23}=10$, and $\tau_c$ can be approximately given by $\ln \frac{0.8}{0.1}/(C_{12}+C_{13}C_{23}/(C_{13}+C_{23})) \approx 0.2079$ and the minimum $\tau_c$ from the numerical solutions is 0.23. Numerically, $\tau_R$ is 0.1221.}\label{taus1}
\end{figure}

\begin{figure}[htb]
\includegraphics[scale=1.2]{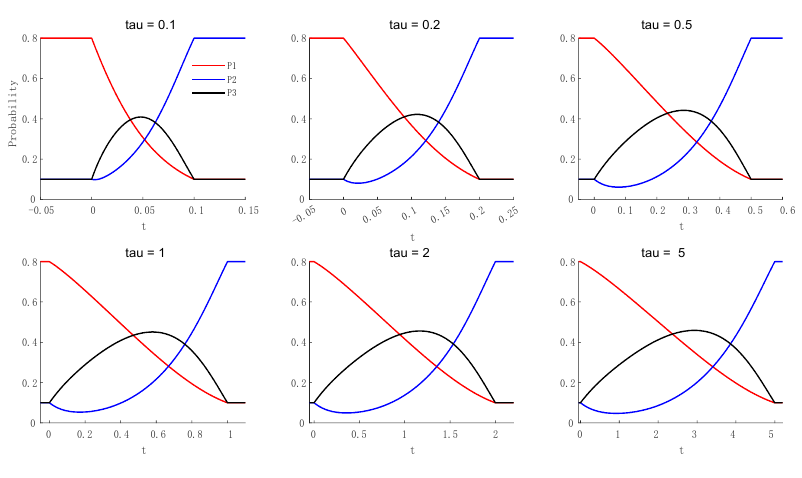}
\caption{Phase II: $C_{12}=5,C_{13}=20,C_{23}=500$, and $\tau_c$ can be approximately given by $\ln \frac{0.8}{0.1}/(C_{12}+C_{13}C_{23}/(C_{13}+C_{23})) \approx 0.0858$ and the minimum $\tau_c$ from the numerical solutions is 0.08319. Numerically, $\tau_R$ is 0.0400.}\label{taus2}
\end{figure}

\begin{figure}[htb]
\includegraphics[scale=1.4]{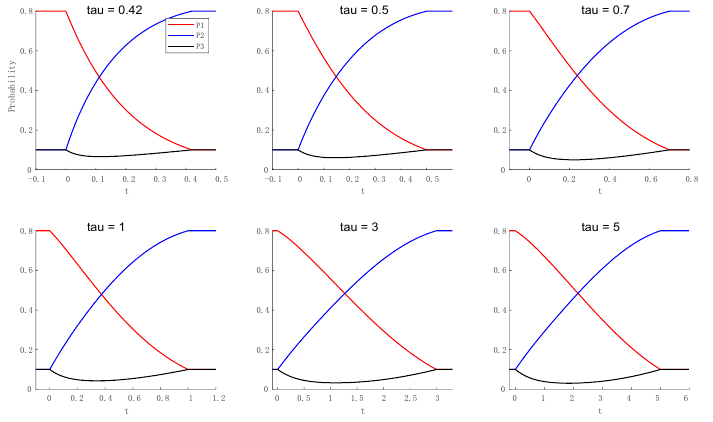}
\caption{Phase III: $C_{12}=5,C_{13}=0.1,C_{23}=50$, where $C_{12}\ll C_{13}C_{23}/(C_{13}+C_{23})$, and $\tau_c$ can be approximately given by $\ln \frac{0.8}{0.1}/(C_{12}+C_{13}C_{23}/(C_{13}+C_{23})) \approx 0.4077$ and the minimum $\tau_c$ from the numerical solutions is 0.409. Numerically, $\tau_R$ is 0.1961.}\label{taus3}
\end{figure}

\begin{figure}[htb]
\includegraphics[scale=1.3]{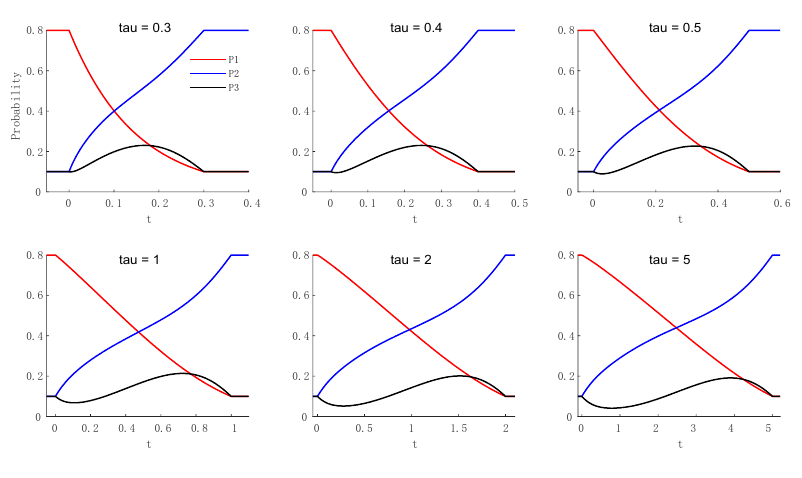}
\caption{Phase IV: $C_{12}=5,C_{13}=2,C_{23}=400$, and $\tau_c$ can be approximately given by $\ln \frac{0.8}{0.1}/(C_{12}+C_{13}C_{23}/(C_{13}+C_{23})) \approx 0.2975$ and the minimum $\tau_c$ from the numerical solutions is 0.29709. Numerically, $\tau_R$ is 0.1429.}\label{taus4}
\end{figure}

We decrease $\tau$ until the optimization problem has no solution by using the OC toolbox and the minimum $\tau$ is regarded as $\tau_c$  approximately. Interestingly, the numerical $\tau_c$ coincides with $\ln \frac{P_1(0)}{P_1(\tau)}/(C_{12}+C_{13}C_{23}/(C_{13}+C_{23}))$. Together with the definition of the averaged relaxation time $\tau_R = \frac{1}{\tau_c}\int_0^{\tau_c} \frac{1}{\lambda^*(t)} dt$, numerically, we show that $\tau_R$ is of the same order as $\tau_c$ in the cases of 4 distinct Phases in Fig.~\ref{taus1}-\ref{taus4}.

For fixed set of channel capacities $C_{ij}$, the optimal  $\textbf{P}^*(t)$ trajectories will change with the control time $\tau$ but restricted in these 4 phases, as shown in Fig.~\ref{taus1}-\ref{taus4}. In other word, as the control time $\tau$ decreases until $\tau_c$, the critical lines partitioning different phases will shift or deform. Consequently, these 4 phases can be deemed as the basic transport phases within the chemical reaction network. Other intricate phases discerned in more extensive networks will invariably be their derivatives.

Basically, the times of flow redirection in channel 3-2 is limited if the total control time $\tau$ becomes shorter, since there will not be enough time for the flow to redirect too many times in a fast time window, otherwise the task of probability transportation can not be complete up to the capacity of the channels. As a consequence, Phase II and IV will transform to the basic Phase I as $\tau$ decreases while Phase I and III change little.

\section{Codes availability}
The toolbox used for numerical solutions of the OC problem is the ``ICLOCS"\footnote{http://www.ee.ic.ac.uk/ICLOCS/default.htm}. The codes used in this work are available at the public repository.

\section{Supplementary figures}


\begin{figure}[htb]
\includegraphics[scale=1]{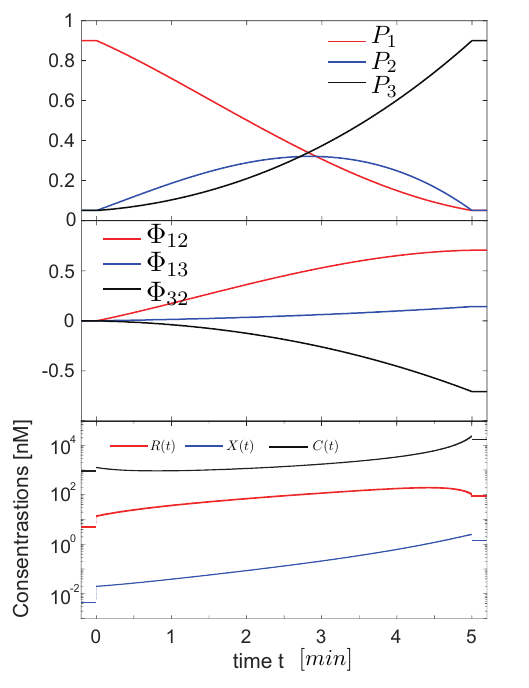}
\caption{The optimal protocols for the 3-state gene regulation network. The numerical solution is obtained via the OC toolbox directly. Driving by the optimal protocol, (a) the upper figure shows that, although state-2 is not the target, its probability accumulates transiently as it serves as a buffer to minimize the total energy dissipation (entropy production) of the switching process due to the resistance $R_{12}=\frac{1}{2k_{-r}}\frac{1}{P_2(t)}$; (b) the middle figure shows that, most of probability flows through channel 1-2-3 rather than the direct pathway 1-3, where $\Phi_{ij}(t) = \int_{0}^{t} J_{ij} dt$ and $\Phi_{12}(\tau) = -\Phi_{32}(\tau) \gg \Phi_{13}(\tau)$; (c) the concentrations $R(t)$, $X(t)$, $C(t)$ all increase and have jumps at the start and the end of the control.}\label{corepressor_big}
\end{figure}

\bibliography{bibliography}